\documentclass[a4paper,11pt,nosumlimits]{amsart}
\usepackage{latexsym,amsmath,amssymb,color,mathrsfs,enumerate,bbm,fullpage,amscd}
\usepackage[english]{babel}
\usepackage[all]{xy}
\usepackage{nicefrac,mathtools}
\usepackage{microtype}

\numberwithin{equation}{section}

\title{The Noncommutative Topology of \\ [2pt] Anti-Self-Dual Gauge Fields}
\author{Simon Brain}
\address{Unit\'e de Recherche en Math\'ematiques, Universit\'e du Luxembourg (Campus Kirchberg), 6 rue Coudenhove--Kalergi, L-1359 Luxembourg, Grand-Duch\'e du Luxembourg}\email{simon.brain@uni.lu}

\date{2nd April 2013}

\newtheorem{thm}{Theorem}[section]

\newtheorem{lem}[thm]{Lemma}
\newtheorem{prop}[thm]{Proposition}
\newtheorem{rem}[thm]{Remark}
\theoremstyle{definition}
\newtheorem{defn}[thm]{Definition}
\newtheorem{example}[thm]{Example}

\renewcommand{\H}{\mathcal{H}}
\newcommand{\Om}{\Omega}
\newcommand{\la}{\langle}
\newcommand{\ra}{\rangle}
\newcommand{\n}{\nabla}
\newcommand{\M}{\textup{M}}
\newcommand{\End}{\textup{End}}
\newcommand{\id}{\textup{id}}
\newcommand{\D}{\textup{d}}
\newcommand{\ii}{\mathrm{i}}
\newcommand{\J}{\textrm{J}}
\newcommand{\hGamma}{\widehat\Gamma}
\newcommand{\h}{\mathcal{H}}
\newcommand{\Dom}{\mathfrak{Dom}}

\DeclareFontFamily{OT1}{pzc}{}
\DeclareFontShape{OT1}{pzc}{m}{it}{<-> s * [1.10] pzcmi7t}{}
\DeclareMathAlphabet{\mathpzc}{OT1}{pzc}{m}{it}

\newcommand{\qp}{{\sf q}}
\newcommand{\sfM}{\sf M}

\newcommand{\A}{\mathcal{A}}
\newcommand{\cB}{\mathcal{B}}
\newcommand{\E}{\mathcal{E}}

\newcommand{\GL}{\textup{GL}}
\newcommand{\U}{\textup{U}}

\newcommand{\B}{\mathbb{B}}
\newcommand{\C}{\mathbb{C}}
\newcommand{\CP}{\mathbb{C}\mathbb{P}}
\newcommand{\HH}{\mathbb{H}}
\newcommand{\RR}{\mathbb{R}}

\newcommand{\TT}{\mathbb{T}}
\newcommand{\ZZ}{\mathbb{Z}}

 \def\lcross{{>\!\!\!\tl}}

\newcommand{\tl}{\triangleleft}

\begin{document}

\begin{abstract}
Through techniques afforded by $C^*$-algebras and Hilbert modules, we study the topology of spaces which parametrize families of instanton gauge fields on noncommutative Euclidean four-spheres $S^4_\sigma$. By deforming the ADHM construction of instantons on the classical sphere $S^4$, we obtain families of instantons on the quantum sphere which are naturally parametrized by noncommutative topological spaces. Using the internal gauge theory of $S^4_\sigma$ determined by the inner automorphisms of its function algebra, we find that one may always recover a classical parameter space by making a suitable choice of internal gauge.
\end{abstract}

\maketitle \tableofcontents

\parskip 1ex

\section{Introduction}
The moduli space of anti-self-dual gauge fields (instantons) on a compact four-dimensional spin manifold encodes a vast amount of information about its differential structure. Although the field of differential geometry is well over a hundred years old, the study of instanton moduli spaces has over the past three decades inspired some great leaps forward in the development of the subject as a modern entity \cite{dk}.

On the other hand, the field of noncommutative differential geometry is still relatively new: our understanding of the differential structure of noncommutative spin manifolds is very much in its infancy. From this point of view, it is only natural to hope that studying moduli spaces of instantons on noncommutative four-manifolds will lead to some insight into the parameters on which their differential structures might depend \cite{ns:ins}. Recent years have witnessed some rapid progress in this direction although, for the most part, this has been purely at the algebraic level \cite{lprs:ncfi,bl:adhm,bl:mod,bvs}. In this paper we continue our somewhat more refined approach to the study of instantons on noncommutative manifolds \cite{blvs}, which takes place at the much richer analytic level of noncommutative topology afforded by $C^*$-algebras and the differential structure of noncommutative spin manifolds encoded by Connes' notion of a spectral triple \cite{ac:book}.

Arguably the simplest example of a four-dimensional noncommutative spin manifold is the quantum four-sphere $S^4_\sigma$ of \cite{cl:id}, obtained via `isospectral deformation'  of the classical sphere $S^4$ along the isometric action of a two-torus $\TT^2$, where $\sigma$ is a deformation parameter. This quantum sphere is therefore a natural point of departure for our exploration of the structure of instanton moduli spaces in noncommutative geometry. 

In \cite{bl:mod} it was shown that, at the level of describing spaces in terms of their coordinate algebras, it is possible to have both classical and noncommutative parameter spaces for the same system of instantons on $S^4_\sigma$. These parameter spaces are related by new `internal' gauge transformations determined by the inner automorphisms of the function algebra 
of the underlying noncommutative manifold \cite{ac:fncg}, which for classical spaces are necessarily trivial. The goal of the present paper is to make these ideas more precise, by showing that they have a very natural interpretation at the more interesting level of $C^*$-algebras and spectral triples.

In passing from algebraic geometry to the more complicated setting of operator spaces, we are led to adopt a much more careful description of the construction of instantons and, in particular, of their gauge theory. In classical geometry, the gauge group of a Hermitian vector bundle $E$ over $S^4$
consists of the set of unitary vertical automorphisms of the differential fibration $E\to S^4$. Over the noncommutative sphere $S^4_\sigma$, the new internal gauge symmetries afforded by the noncommutativity of the function algebra $C(S^4_\sigma)$ must also be incorporated into the group of gauge symmetries \cite{ac:fncg,bmvs}. 

The paper is organized as follows. In \S\ref{se:cpas} we sketch Kasprzak's recent approach \cite{kasp} to Rieffel's strict deformation quantization \cite{mr:def}; this is the deformation procedure that we shall use herein to obtain noncommutative spaces from classical ones. The strategy is built upon Landstad's beautiful characterization of crossed product $C^*$-algebras \cite{land}, which uses the Takai-Takesaki dual action to determine the conditions under which a given $C^*$-algebra $B$ is isomorphic to a crossed product $B\simeq A\lcross\,\Gamma$ for a given Abelian group $\Gamma$. Kasprzak's observation is that this dual action may be twisted using a given choice of two-cocycle $\sigma$ on the Pontryagin dual group $\hGamma$ to obtain a new dual action upon $B$ and hence a different but isomorphic factorization of the crossed product $B\simeq A^\sigma\lcross\,\Gamma$. In contrast to Rieffel's original approach to strict deformation quantization (which goes first via group actions on smooth subalgebras), Kasprzak's version has the advantage of being applicable directly at the $C^*$-algebraic level and is therefore particularly appropriate to the techniques of noncommutative topology used in the present paper.

In \S\ref{se:twistors} we apply the Landstad--Kasprzak theory of twisted $C^*$-algebras to obtain the noncommutative sphere $S^4_\sigma$ from its classical counterpart $S^4$. We then extend this to obtain a noncommutative analogue of the Penrose twistor fibration $\CP^3\to S^4$, the most important geometric ingredient in the construction of instantons on the four-sphere. Indeed, every instanton bundle over $S^4$ is equivalent via pull back along the twistor fibration to a holomorphic vector bundle over twistor space $\CP^3$. By pulling them back to the homogeneous twistor space $\C^4\to \CP^3$, such bundles are in turn quite easily constructed using algebraic methods \cite{ma:gymf}. The approach of \cite{bl:adhm,bl:mod,bvs} was to show that the same is true of instanton bundles over the noncommutative four-sphere; we adopt the same strategy herein. 

Of particular novel value is our passage between the coordinate-algebraic description of homogeneous twistor space and its topological description at the $C^*$-algebraic level. Since this is a construction involving non-compact spaces, this is problematic even in the classical case (where coordinate functions act as unbounded operators on the space of continuous functions). Our approach to this problem for the noncommutative twistor spaces constructed in the present paper is essentially an adaptation of techniques appearing in \cite{kasp}, although it perhaps suggests a new approach to a more general reconciliation of algebraic- and differential-geometric techniques in the noncommutative setting.

In \S\ref{se:gauge theory} we recall the basic definitions of Connes' theory of noncommutative spin manifolds, describing in particular how the Landstad--Kasprzak theory extends to give a procedure which yields isospectral deformations of spectral triples over classical spin manifolds. We then recall the gauge theory of a noncommutative spin manifold as described in \cite{ac:fncg,bl:adhm,bl:mod} and elaborated upon in \cite{bmvs}. 

Finally in \S\ref{se:final} we present the main results of the paper. Using our description of the noncommutative twistor fibration, we present the construction of instantons on the classical sphere $S^4$ from the point of view of $C^*$-algebras and noncommutative topology, which we then deform using the Landstad--Kasprzak deformation theory. The parameter spaces of instantons so obtained are {\em a priori} noncommutative: we show that, using the internal gauge theory of the noncommutative four-sphere, one may always recover a classical space of parameters. In conclusion we deduce that the gauge theory of the quantum four-sphere is an important part of its noncommutative geometry, in the sense that the topology of the space of instanton gauge fields on $S^4_\sigma$ tells us a great deal about its differential structure.

\subsubsection*{Some basic notation and terminology}\label{se:aff}  Given a separable $C^*$-algebra $A$, we write $\M(A)$ for the multiplier algebra of $A$, usually equipped with the strict topology. Recall that a $C^*$-algebra $A$ is unital if and only if $A=\M(A)$. We write $\mathrm{Aut}(A)$ for the group of all $*$-automorphisms of $A$. By a morphism of $C^*$-algebras $A\to B$
we mean a continuous $*$-algebra map $\phi:A\to \M(B)$ from $A$ into the multiplier
algebra $\M(B)$ such that $\phi(A)B$ is norm dense in $B$. 

We describe unbounded linear operators on $A$ using the notation $T:\Dom(T)\to A$, where $\Dom(T)\subseteq A$ denotes the (necessarily dense) domain of $T$. Let $T:\Dom(T)\to A$ be an unbounded linear operator on $A$. Then $T$ is said to be {\em affiliated} to $A$, and we write $T\eta A$, if there exists an element $\mathfrak{z}_T\in \M(A)$ such that $||\mathfrak{z}_T||\leq 1$ and
$$
\left( x\in \Dom(T)~\text{and}~y=Tx\right)~ \Leftrightarrow ~\left( \exists\, a\in A ~\text{such that}~ x=(1-(\mathfrak{z}_{_T})^*\mathfrak{z}_{_T})^{1/2}a ~\text{and}~y=\mathfrak{z}_Ta\right).
$$
We write $A^\eta$ for the set of all linear operators affiliated to $A$. The element $\mathfrak{z}_T$ is called the {\em $\mathfrak{z}$-transform} of $T$. It has the property that a multiplier $z\in \M(A)$ is the $\mathfrak{z}$-transform of some affiliated element $T\in A^\eta$ if and only if $||z||\leq 1$ and $(1-z^*z)^{1/2}A$ is linearly dense in $A$.
Amongst the many useful properties of affiliated elements, one finds that $\M(A)\subseteq A^\eta$ and that if $T\in A^\eta$ and $a\in \M(A)$ then both $Ta\in A^\eta$ and $aT\in A^\eta$. In particular, if $A$ is a unital $C^*$-algebra, then $A^\eta=A$. 

For an arbitrary $C^*$-algebra $A$ there is in general no way to add or multiply affiliated elements. However, when $A=C_0(X)$ is a commutative $C^*$-algebra one finds that $A^\eta=C(X)$, the algebra of continuous functions on $X$. Indeed, if $a\in C(X)$, let $T$ be the operator on $A$ given by multiplication by $a$. By definition, we have 
$$
\Dom(T):=\{ f\in C_0(X)~|~\textup{lim}_{x\to\infty}a(x)f(x)=0\}.
$$ 
One checks that the $\mathfrak{z}$-transform of $T$ is the function $\mathfrak{z}_T(x):=a(x)\left(1+a^*(x)a(x)\right)^{-1/2}$ and hence that $T\in A^\eta$.
Conversely, if $T\in A^\eta$ then $\mathfrak{z}_T$ belongs to the multiplier algebra $\M(A)=C_b(X)$, with $||\mathfrak{z}_T||\leq 1$ and $|\mathfrak{z}_T(x)|<1$ for all $x\in X$ (otherwise the set $\Dom(T)=(1-\mathfrak{z}_T^*\mathfrak{z}_T)^{1/2}A$ would not be dense in $A$). It follows that, by setting 
$$
a(x):=\mathfrak{z}_T(x)\left( 1-\mathfrak{z}_T^*(x)\mathfrak{z}_T(x)\right)^{-1/2},
$$
we obtain an element $a\in C(X)$ such that the operator $T$ coincides with multiplication by $a$. We shall use the affiliation relation extensively in the present paper, as a way of relating elements of the function spaces $C(X)$ and $C_0(X)$ (and their generalizations to the noncommutative setting).

\subsubsection*{Acknowledgements} This project was supported by the National Research Fund, Luxembourg, and cofunded under the Marie Curie Actions of the European Commission (FP7-COFUND). The author thanks Pawe\l\  Kasprzak, Bram Mesland and Walter van Suijlekom for helpful conversations and an anonymous referee for some very useful comments which corrected a number of mistakes in the paper. It is a pleasure to thank the organizers of the workshop on `Noncommutative Algebraic Geometry and its Applications to Physics' which took place at the Lorentz Centre in Leiden, Netherlands, March 2012, where some of the ideas presented herein were developed.

\section{Duality for Crossed Product $C^*$-Algebras}\label{se:cpas}
In the present paper the fundamental construction will be that of the
crossed product algebra $A\lcross_\alpha\,\Gamma$ associated to the action $\alpha:\Gamma\to\textrm{Aut}(A)$
of a locally compact Abelian group $\Gamma$ on a $C^*$-algebra $A$.
In this section we review the various aspects of the theory that we
shall need.

\subsection{Landstad duality for covariant systems}\label{se:land}
In this paper, by a {\em covariant system} we shall mean a triple
$(A,\Gamma,\alpha)$, where $\Gamma$ is a locally compact Abelian
group and $A$ is a $C^*$-algebra equipped with a strongly continuous
action $\alpha:\Gamma\to\mathrm{Aut}(A)$. Associated to the system
$(A,\Gamma,\alpha)$ there is the $*$-algebra $C_c(\Gamma,A)$ of
continuous compactly supported $A$-valued functions on $\Gamma$,
equipped with the multiplication and involution operations
\begin{equation}\label{conv}
(f_1\star f_2)(g)=\int_\Gamma f_1(r)\,\alpha_r(f_2(r^{-1}g))\D
r,\qquad f^*(g)=f(g^{-1})^*,
\end{equation}
for each $f_1,f_2,f\in C_c(\Gamma,A)$ and $g\in\Gamma$, where the integral is with respect to the Haar measure on $\Gamma$.
The {\em crossed product algebra} associated to the system
$(A,\Gamma,\alpha)$ is the completion of $C_c(\Gamma,A)$ with
respect to the $C^*$-norm induced by the left regular representation; we denote it by
$A\lcross_\alpha\,\Gamma$.

\begin{rem}\textup{
The crossed product algebra associated to the covariant system $(\C,\Gamma,\id)$ is nothing other than the group $C^*$-algebra $C^*(\Gamma)$, i.e. the completion of the $*$-algebra of continuous compactly supported functions $C_c(\Gamma)$ equipped with the usual convolution product and involution.
}
\end{rem}

A {\em covariant representation} of the system $(A,\Gamma,\alpha)$
is a pair $\mu=(\mu_\Gamma,\mu_A)$, where
$\mu_\Gamma:\Gamma\to\U(\h_\mu)$ is a unitary group representation
and $\mu_A:A\to \B(\h_\mu)$ is a representation of $A$ by bounded
operators on the same Hilbert space $\h_\mu$, which together obey
the {\em covariance condition}
\begin{equation}\label{covrep}
\mu_\Gamma(g)\mu_A(a)\mu_\Gamma(g)^*=\mu_A(\alpha_g(a))
\end{equation}
for all $g\in \Gamma$, $a\in A$.  The representation theory of the
crossed product algebra $A\lcross_\alpha\,\Gamma$ completely encodes
the covariant representations of the triple $(A,\Gamma,\alpha)$.
Indeed, every covariant representation of $(A,\Gamma,\alpha)$ on
$\h_\mu$ gives rise to a representation of $A\lcross_\alpha\,\Gamma$
on $\h_\mu$ via the integrated form
$$
\mu(f):=\int_\Gamma \mu_A(f(r))\mu_\Gamma(r)\D r, \qquad f\in C_c(\Gamma,A),~r\in \Gamma.
$$
Conversely, given a representation $\mu:A\lcross_\alpha\,\Gamma\to \B(\h_\mu)$, one uses the canonical inclusions 
\begin{equation}\label{canincs}
\iota_\Gamma:\Gamma\to\M(A\lcross_\alpha\,\Gamma),\qquad\iota_A:A\to\M(A\lcross_\alpha\,\Gamma),
\end{equation}
to recover a covariant representation $(\mu_\Gamma,\mu_A)$ of $(A,\Gamma,\alpha)$ by setting $\mu_\Gamma:=\mu\circ\iota_\Gamma$ and $\mu_a:=\mu\circ\iota_A$.

Next we come to describe Landstad duality for covariant systems
\cite{land}. Given a locally compact Abelian group $\Gamma$, we write $\hGamma$ for its
Pontryagin dual group. The fundamental notion in Landstad's theory is that of
a {\em $\Gamma$-product}, whose definition we now recall.

\begin{defn}\label{de:g-prod}
A $C^*$-algebra $B$ is said to be a {\em $\Gamma$-product} if:
\begin{enumerate}[\hspace{0.5cm}(i)]
\item there is a continuous homomorphism
$$
\lambda:\Gamma\to \U \M(B), \qquad g\in \Gamma\mapsto \lambda_g,
$$
of $\Gamma$ into the unitary group of $\M(B)$;
\item there is a homomorphism
$$
\hat\alpha:\hGamma\to\mathrm{Aut}(B), \qquad \xi\in
\hGamma\mapsto\hat\alpha_\xi\in \mathrm{Aut}(B),
$$
such that $(B,\hGamma,\hat\alpha)$ is a covariant system and
$\hat\alpha_\xi(\lambda_g)=\xi(g) \lambda_g$ for all $g\in \Gamma$
and $\xi\in \hGamma$.
\end{enumerate}
\end{defn}

By integration, the unitary representation $\lambda:\Gamma\to \U
\M(B)$ extends to a $*$-algebra map $C_c(\Gamma)\to \M(B)$ and hence
to a morphism of $C^*$-algebras $C^*(\Gamma)\to B$. Identifying
$C^*(\Gamma)$ with $C_0(\hGamma)$ via Fourier transform,
we get an injective morphism $\lambda:C_0(\hGamma)\to B$, realizing
$C_0(\hGamma)$ as a subalgebra of $\M(B)$. Given a $\Gamma$-product
$(B,\lambda,\hat\alpha)$, there is an important role attached to the
set of elements of the multiplier algebra $\M(B)$ obeying the
following special conditions.

\begin{defn}\label{de:l-cons}
Let $B$ be a $\Gamma$-product. An element $x\in \M(B)$ is said
to satisfy {\em Landstad's conditions} if:
\begin{enumerate}[\hspace{0.5cm}(i)]
\item $x$ is a fixed point of the $\hGamma$-action, namely $\hat\alpha_\xi(x)=x$ for all $\xi\in\hGamma$;
\item for all $g\in \Gamma$ the map $g\mapsto \lambda_g x\lambda_g^*$
is norm continuous;
\item for all $f_1,f_2\in C_0(\hGamma)$ we have $\lambda_{f_1} x\lambda_{f_2}\in B$.
\end{enumerate}
\end{defn}

In terms of these definitions, the essence of Landstad duality is
the following. Given a covariant system $(A,\Gamma,\alpha)$, there
is an action $\hat\alpha$ of $\hGamma$ on $C_c(\Gamma,A)$ defined by
$$
(\hat\alpha_\xi(f))(g):=\xi(g)f(g),\qquad f \in C_c(\Gamma,A),~ g\in
\Gamma,
$$
which extends to an action
$\hat\alpha:\hGamma\to\mathrm{Aut}(A\lcross_\alpha\,\Gamma)$ such
that the triple $(A\lcross_\alpha\,\Gamma,\hGamma,\hat\alpha)$ is a
covariant system. Moreover, the $C^*$-algebra
$A\lcross_\alpha\,\Gamma$ is a $\Gamma$-product and each element in
$A$ satisfies Landstad's conditions. The triple
$(A\lcross_\alpha\,\Gamma,\hGamma,\hat\alpha)$ is called the {\em
dual system} of $(A,\Gamma,\alpha)$. The action $\hat\alpha$ of
$\hGamma$ is nothing other than the Takai-Takesaki dual action; it has the
property that, modulo compact operators, the `double dual' system
$(A\lcross_\alpha\, \Gamma)\lcross_{\hat\alpha}\, \hGamma$ is
isomorphic to $A$.

Conversely, a $C^*$-algebra $B$ is a $\Gamma$-product for a given
Abelian group $\Gamma$ only if there is a covariant system
$(A,\Gamma,\alpha)$ such that $B=A\lcross_\alpha\,\Gamma$. In
particular, the $C^*$-algebra $A$ consists of those elements in
$\M(B)$ satisfying Landstad's conditions, whereas the $\Gamma$-action
is defined by $\alpha_g(a):=\lambda_g a \lambda_g^*$ for each $a\in
A$ and $g\in \Gamma$. The triple $(A,\Gamma,\alpha)$ is unique up to
isomorphism.

In the special case where $\Gamma$ is a compact group, its Pontryagin dual group $\hGamma$ is discrete.
For each $\zeta\in\hGamma$, we define the corresponding {\em $\zeta$-spectral subspace} by
$$
A_\zeta:=\{a\in A~|~\alpha_r(a)=\zeta(r) a~\text{for all}~ r\in\Gamma\}.
$$ 
These spectral subspaces give a $\hGamma$-grading of $A$, in the sense that there is a decomposition
\begin{equation}\label{specde}
A=\bigoplus_{\zeta\in\hGamma} \,A_\zeta,
\end{equation}
with infinite series on the right hand side converging on an appropriate dense subalgebra of $A$ \cite{quigg}. 

\subsection{Landstad-Kasprzak deformation theory}\label{se:kasp}
Landstad duality identifies the conditions under which
a given $C^*$-algebra $B$ can be decomposed as a crossed product
$B=A\lcross_\alpha\Gamma$ for a given Abelian group $\Gamma$, the
crucial component of this characterization being the Takai-Takesaki dual
action $\hat\alpha$. Kasprzak's observation \cite{kasp} is that the
action $\hat\alpha$ may be deformed to give a new
action and consequently a new $\Gamma$-product
structure on $B$. 

By a {\em two-cocycle} $\sigma$ on $\hGamma$ we mean
a continuous map $\sigma:\hGamma\times \hGamma\to \mathbb{T}$ with
values in the circle group $\mathbb{T}$ obeying the identities
\begin{align*}
&\sigma(e,\xi)=\sigma(\xi,e)=1,&
\sigma(\xi,\eta)\sigma(\xi+\eta,\zeta)=\sigma(\xi,\eta+\zeta)\sigma(\eta,\zeta)&
\end{align*}
for all $\xi,\eta,\zeta\in\hGamma$.

\begin{rem}\label{re:bic}\textup{A {\em bicharacter} on the group $\hGamma$ is a continuous map $\sigma:\hGamma\times \hGamma\to \mathbb{T}$ such that the maps $\sigma^1_\xi: \eta\mapsto \sigma(\xi,\eta)$
and $\sigma^2_\eta:\xi\mapsto \sigma(\xi,\eta)$ each define
characters of the group $\hGamma$ and hence elements of the group
$\Gamma$. Since $\hGamma$ is Abelian, every two-cocycle on $\hGamma$ is cohomologous to a bicharacter \cite{klep}.}
\end{rem}

Given a $\Gamma$-product $(B,\lambda,\hat\alpha)$ and a two-cocycle
$\sigma$ on $\hGamma$, for each $\xi\in\hGamma$ we define a unitary element $U_\xi:=\lambda(\sigma^1_\xi)$ of the multiplier algebra $\M(B)$. Then there is a twisted action of
$\hGamma$ on $B$ defined by
\begin{equation}\label{kasp}
\hat\alpha^\sigma:\hGamma\to\textup{Aut}(B),\qquad \hat\alpha^\sigma_\xi(b):=U_\xi^*\,\hat\alpha_\xi(b)\,U_\xi,
\end{equation}
for each $b\in B$, $\xi\in\hGamma$.
This action too obeys the relation
$$
\hat\alpha^\sigma_\xi(\lambda_g)=\xi(g)\,\lambda_g,\qquad
\xi\in\hGamma,~g\in\Gamma,
$$
and hence it defines a $\Gamma$-product structure on the
$C^*$-algebra $B$. Immediately from the Landstad theory it follows
that there must exist a $C^*$-algebra $A^\sigma$ and an action $\alpha^\sigma:\Gamma\to \mathrm{Aut}(A^\sigma)$
such that $(A^\sigma,\Gamma,\alpha^\sigma)$ is a covariant system
and $B= A^\sigma\lcross_{\alpha^\sigma}\,\Gamma$.

\begin{rem}\textup{
One of the main results in \cite{kasp} is that, if $\sigma_1$ and $\sigma_2$ are cohomologous cocycles, then the corresponding $C^*$-algebras $A^{\sigma_1}$ and $A^{\sigma_2}$ are canonically isomorphic. In light of Rem.~\ref{re:bic}, we may as well work exclusively with bicharacters instead of two-cycles.
}
\end{rem}

The $C^*$-algebra $A^\sigma$ is identified with the fixed points in
$\M(B)$ under the action $\hat\alpha^\sigma$ of $\hGamma$. It carries
the $\Gamma$-action $\alpha^\sigma$ defined by
$$
\alpha^\sigma:\Gamma\to\textup{Aut}(A^\sigma),\qquad \alpha^\sigma_g(a)=\lambda_g a \lambda_g^*,
$$
for each $a\in A^\sigma$, $g\in\Gamma$.
Note that it is not the formula defining the $\Gamma$-action which
changes under the deformation, rather its domain of definition.

Following \cite{hanmat}, we can be rather more precise about the
algebras $A$ and $A^\sigma$. Indeed, $A$ is by construction identified with the fixed
subalgebra of $\M(A\lcross_\alpha\,\Gamma)$ under the $\hGamma$-action defined by
$\hat\alpha_\xi(f)(g)=\xi(g)f(g)$, where $\xi\in\hGamma$, $g\in
\Gamma$ and $f\in C_c(\Gamma,A)$. The fixed points of this action are precisely the set of
distributions concentrated at the group identity of $\Gamma$, which
make sense as elements of the multiplier algebra. They give an
algebra isomorphic to $A$.

On the other hand, for the algebra $A^\sigma$ the dual action is
defined using the elements $U_\xi:=\lambda (\sigma^1_\xi)$ for $\xi\in\hGamma$. By the
covariance property \eqref{covrep}, the adjoint action of $U_\xi$
coincides with the action of $\alpha_{\sigma^1_\xi}$ and so it
follows that
$$
(\hat\alpha^\sigma_\xi(f))(g)=\alpha^{-1}_{\sigma^1_\xi}\left(\hat\alpha_\xi(f)(g)\right)=\xi(g)\,\alpha^{-1}_{\sigma^1_\xi}(f(g))
$$
for all $g\in\Gamma$, $\xi\in \hGamma$ and $f\in C_c(\Gamma,A)$.
The fixed subalgebra where $\hat\alpha^\sigma_\xi(f)=f$ is therefore
seen to consist of those functions $f$ satisfying
$$
\alpha_{\sigma^1_\xi}(f(g))=\xi(g)f(g), \text{for all} ~g\in \Gamma,~\xi\in \hGamma,
$$
for all  $g\in \Gamma$, $\xi\in \hGamma$, upon making a suitable change of variable.

As one might expect, this deformation construction has a number of important
functorial properties \cite{kasp}. In particular, let
$(I,\Gamma,\alpha_{_I})$, $(A,\Gamma,\alpha)$ and $(B,\Gamma,\beta)$
be covariant systems and suppose there is an exact sequence of
$C^*$-algebras
$$
0\to I \to A\xrightarrow{\pi}B\to 0
$$
consisting of $\Gamma$-equivariant morphisms. Let $\sigma$ be a
two-cocycle on $\hGamma$ and let $I^\sigma$, $A^\sigma$ and
$B^\sigma$ be the Landstad $C^*$-algebras associated to the
deformed dual actions.  
Then there is a $\Gamma$-equivariant exact
sequence of $C^*$-algebras
$$
0\to I^\sigma \to A^\sigma\xrightarrow{\pi^\sigma} B^\sigma\to 0,
$$
where the morphism $\pi^\sigma:A^\sigma\to B^\sigma$ is the
restriction of the morphism $\pi:A\lcross_\alpha\,\Gamma\to
B\lcross_{\beta}\,\Gamma$ to the Landstad algebra
$A^\sigma\subseteq \M(A\lcross_\alpha\,\Gamma)$. The deformation operation $A\mapsto A^\sigma$ therefore gives an exact covariant functor from
the category of $\Gamma$-equivariant $C^*$-algebras to itself.

\section{The Noncommutative Twistor Fibration}\label{se:twistors}

As already mentioned, one of the crucial ingredients in the study of instantons on the classical sphere $S^4$ is the Penrose twistor fibration $\CP^3\to S^4$. It encapsulates in its geometry the very nature of the anti-self-duality equations, enabling us to reinterpret instanton bundles on $S^4$ via pull-back in terms of algebraic vector bundles over $\CP^3$. In this section we briefly recall the details of this fibration from a topological point of view, which we then deform using the Landstad-Kasprzak deformation theory.

\subsection{Classical and noncommutative twistor theory}\label{se:def-tw}
In this section we recall the coordinate-algebraic description of the various spaces which constitute the twistor fibration, as a way to understand the maps involved \cite{bm:qtt}. Then by passing to the level of $C^*$-algebras, we apply the deformation theory of the previous section to obtain a noncommutative analogue of the twistor space construction. We begin with the four-dimensional sphere $S^4$.

\begin{defn}
The algebra $\A[S^4]$ of coordinate functions on the four-sphere is the commutative unital $*$-algebra generated by the complex coordinate functions
$x_1$, $x_2$, their conjugates $x_1^*$, $x_2^*$ and the real coordinate function $x_0=x_0^*$, subject to the sphere relation
\begin{equation}\label{four-sph}
x_1^*x_1+x_2^*x_2+x_0^2=1.
\end{equation}
\end{defn}

To study instantons on the Euclidean space $S^4$, one needs two versions of its associated twistor space. The first of these is nothing other than the complex vector space $\C^4$, which we refer to as {\em homogeneous twistor space}. The second is its corresponding projectivization $\CP^3$, which we refer to as {\em projective twistor space}.

\begin{defn} The algebra $\A[\C^4]$ of polynomial functions on homogeneous twistor space $\C^4$ is the commutative unital $*$-algebra generated by the coordinate functions $z_j$, $j=1,\ldots,4$, together with their conjugates $z_l^*$, $l=1,\ldots,4$. 
\end{defn}

To obtain a coordinate-algebraic description of projective twistor space $\CP^3$ (thought of as a real manifold), we note that specifying a one-dimensional subspace of $\C^4$ is equivalent to specifying a matrix $e\in\M_4(\C)$ obeying
$$
e^2=e,\qquad e^*=e,\qquad \textup{Tr}\,e=1.
$$
Since $e$ is Hermitian, its eigenvalues must be real. As $e$ is idempotent, these eigenvalues must be elements of the set $\{0,1\}$ and, from the trace condition on $e$, its image must therefore be a line in $\C^4$, i.e. the eigenspace with eigenvalue $1$. This leads to the following definition.

\begin{defn}
The algebra $\A[\CP^3]$ of coordinate functions on projective twistor space $\CP^3$ is the commutative unital $*$-algebra generated by the entries of the self-conjugate matrix
\begin{equation}\label{q}
\qp:=\begin{pmatrix}a_1 & u_1 & u_2 & u_3\\ u_1^* & a_2 & v_3 & v_2 \\
u_2^* & v_3^* & a_3 & v_1 \\ u_3^* & v_2^* & v_1^* &
a_4\end{pmatrix} ,
\end{equation}
subject to the relation $\textup{Tr}\,\qp=1$ and the relations coming from the projection condition
$\qp^2=\qp$, that is to say $\sum_r \qp_{jr}\qp_{rl}=\qp_{jl}$ for each $j,l=1,\ldots,4$. 
\end{defn}

Let us write $\widetilde\C^4:=\C^4\backslash\{0\}$. Then twistor space $\CP^3$ is of course defined by the canonical projection $\widetilde\C^4\to\CP^3$. The coordinate algebra $\A[\widetilde\C^4]$ is by definition obtained by adjoining to $\A[\C^4]$ the inverse of the radius element $\sum_i |z_i|^2$. The fact that $\CP^3$ is just the projectivization of $\C^4$ translates into the fact that there is an injective $*$-algebra map 
\begin{equation}\label{tw-inc}
\eta:\A[\CP^3]\to \A[\widetilde\C^4],\qquad \eta(\qp_{jl}):=\frac{1}{\sum_i |z_i|^2}z_jz_l^*, \qquad j,l=1,\ldots,4.
\end{equation} 
In this way, we immediately find the following coordinate-algebraic description of the Penrose fibration.

\begin{lem}\label{le:pen}
There is a $*$-homomorphism of unital $*$-algebras $\A[S^4]\rightarrow\A[\CP^3]$.
\end{lem}

\proof The required inclusion is given at the level of generators by
\begin{align}\label{hopf}
x_1\mapsto 2(u_2+v_2^*)&=2(z_1z_3^*+z_2^*z_4), \qquad x_2\mapsto 2(v_3-u_3^*)=2(z_2^*z_3-z_1^*z_4), \\
x_0&\mapsto 2(a_1+a_2-1)=z_1^*z_1+z_2^*z_2-z_3^*z_3-z_4^*z_4
\end{align}
and extended as a $*$-algebra map. It is straightforward to check as in \cite{bm:qtt} that the sphere relation in $\A[S^4]$ is equivalent to the trace relation in the algebra $\A[\CP^3]$.
\endproof

We shall need some additional structure on the algebra map $\A[S^4]\rightarrow\A[\CP^3]$. For later use, we introduce the map $\J:\A[\C^4]\to\A[\C^4]$ defined
on generators by
\begin{equation}\label{J}
\J(z_1,z_2,z_3,z_4):=(-z_2^*,z_1^*,-z_4^*,z_3^*)
\end{equation}
and extended as a $*$-algebra map. 
Equipping the algebra $\A[\C^4]$ with the map $\J$ identifies the underlying space $\C^4$ with
the quaternionic vector space $\HH^2$ \cite{lprs:ncfi}. Using the identification of generators \eqref{tw-inc}, the map $\J$
also defines an automorphism of the algebra $\A[\CP^3]$, given on generators by
\begin{align*}
\J(a_1)&=a_2, & \J(a_2)&=a_1, & \J(a_3)&=a_4, & \J(a_4)&=a_3, &
\J(u_1)&=-u_1, \\ \J(v_1)&=-v_1, & \J(u_2)&=v_2^*, &
\J(u_3)&=-v_3^*, & \J(v_2)&=u_2^*, & \J(v_3)&=-u_3^*
\end{align*}
and extended as a $*$-algebra map.

\begin{lem}
The invariant subalgebra of $\A[\CP^3]$ under the automorphism 
$$\textup{J}:\A[\CP^3]\to\A[\CP^3]$$ is isomorphic to the coordinate algebra $\A[S^4]$ of the four-sphere.
\end{lem}

\proof This follows immediately by direct calculation as in \cite{bm:qtt}, using the identification \eqref{hopf} of generators of the algebras $\A[S^4]$ and $\A[\CP^3]$.
\endproof

Having dealt with the algebraic structure of the twistor fibration, we pass to the topological level of $C^*$-algebras. The universal completions of the commutative unital $*$-algebras $\A[S^4]$ and $\A[\CP^3]$ are respectively the unital $C^*$-algebras $C(S^4)$ and $C(\CP^3)$ of
continuous functions on the compact spaces $S^4$ and $\CP^3$. On the other hand, the topology of the locally compact space $\C^4$ is encoded by the non-unital $C^*$-algebra $C_0(\C^4)$ of continuous functions on $\C^4$ vanishing at infinity. 

The following lemma gives a $C^*$-algebraic description of the topology of the twistor fibration, both at the homogeneous and non-homogeneous levels. Recall the definition of a morphism of $C^*$-algebras given in \S\ref{se:aff}.

\begin{lem}
The $*$-algebra map $\A[S^4]\hookrightarrow\A[\CP^3]$ extends to an injective morphism of unital $C^*$-algebras $C(S^4)\to C(\CP^3)$. There is a sequence  
\begin{equation}\label{cl-seq}
C(S^4) \rightarrow C(\CP^3)\to C_0(\widetilde\C^4)
\end{equation}
of injective morphisms of $C^*$-algebras.
\end{lem}

\proof This is a consequence of Gelfand duality, with injectivity corresponding to the fact that the underlying maps are surjective morphisms of topological spaces. The second map is constructed by pulling back elements of $C(\CP^3)$ to obtain elements of the $C^*$-algebra $C_b(\widetilde\C^4)$ of bounded continuous functions on $\widetilde\C^4$. The latter is nothing other than the multiplier algebra $\M(C_0(\widetilde\C^4))$, whence we obtain a morphism $C(\CP^3)\to C_0(\widetilde\C^4)$.\endproof

We are now ready to illustrate the Landstad-Kasprzak deformation theory.
Let us assume that $\Gamma$ is a compact Abelian group acting continuously and equivariantly on the twistor fibration.
By pulling back the actions of $\Gamma$ on these spaces to continuous functions, we find that \eqref{cl-seq} is a $\Gamma$-equivariant sequence of $C^*$-algebras, yielding a corresponding sequence of $\Gamma$-covariant systems.

\begin{prop} 
Given a two-cocycle $\sigma:\hGamma\times\hGamma\to \TT$
there is sequence of injective morphisms
$$
C(S^4_\sigma)\to C(\CP^3_\sigma)\to C_0(\widetilde\C^4_\sigma).
$$ 
\end{prop}

\proof The deformation
theory immediately gives a noncommutative version of the twistor fibration, obtained as application of the following three
steps:
(i) construct the crossed product algebras associated to the above covariant systems; 
(ii) introduce the $\Gamma$-product structures corresponding to the twisted
$\hGamma$-actions;
(iii) let $C(S^4_\sigma)$, $C(\CP^3_\sigma)$ and $C_0(\widetilde\C^4_\sigma)$ be the Landstad
algebras for these new $\Gamma$-products.
Functoriality of the deformation induces the required morphisms.\endproof

As usual in noncommutative topology, we interpret $C(S^4_\sigma)$ and
$C(\CP^3_\sigma)$ as the $C^*$-algebras of continuous
functions on underlying `virtual' spaces $S^4_\sigma$ and
$\CP^3_\sigma$. We think of the morphism $C(S^4_\sigma)\to
C(\CP^3_\sigma)$ as giving a noncommutative
analogue of the Penrose twistor fibration. Similarly, there is a virtual space $\C^4_\sigma$ associated to the deformed $C^*$-algebra $C_0(\C^4_\sigma)$ giving a noncommutative analogue of the homogeneous twistor space. Since the $C^*$-algebra $C_0(\widetilde\C^4)$ is obtained from $C_0(\C^4)$ by a quotient construction (given by taking the dual of the continuous inclusion $\widetilde \C^4\hookrightarrow \C^4$), we also find by functoriality that $C_0(\widetilde\C^4_\sigma)$ is simply a quotient of $C_0(\C^4_\sigma)$ by an appropriate ideal.

\begin{prop}
There is a $\Gamma$-equivariant isomorphism of $C^*$-algebras
$$
\textup{J}_\sigma:C(\CP^3_\sigma)\to C(\CP^3_\sigma)
$$
for which the $C^*$-subalgebra of invariant elements is isomorphic to $C(S^4_\sigma)$.
\end{prop}

\proof It is clear that the map \eqref{J} extends to an isomorphism of $C^*$-algebras $$\J:C(\CP^3)\to C(\CP^3)$$ whose subalgebra of invariant elements is isomorphic to $C(S^4)$. The above assumption that $\Gamma$ acts equivariantly upon the twistor fibration means precisely that $\J$ commutes with the group action on $C(\CP^3)$. Functoriality ensures that, after deformation, we obtain a $\Gamma$-equivariant map with the necessary properties. 
\endproof

\subsection{Algebraic structure of noncommutative twistor space} The previous section described how to deform the twistor fibration at the topological level of $C^*$-algebras. For later use, we shall also need a description of the noncommutative twistor space $\C^4_\sigma$ at the algebraic level, in the sense that we need to understand how to deform the coordinate algebra $\A[\C^4]$ in way which is compatible with the deformation of its topological version $A:=C_0(\C^4)$. 

To this end, we note that $\A[\C^4]$ is a unital $*$-subalgebra of the algebra $C(\C^4)$ of elements affiliated to the $C^*$-algebra $C_0(\C^4)$. Our strategy will be to deform the algebra $\A[\C^4]$ in such a way that the elements of the resulting algebra $\A[\C^4_\sigma]$ are affiliated to the deformed $C^*$-algebra $A^\sigma:=C_0(\C^4_\sigma)$.  

\begin{rem}\label{re:tt}\textup{
The way to proceed now depends upon the group $\Gamma$ and so we have to make a choice. In keeping with the literature \cite{cl:id} we shall assume henceforth that $\Gamma=\TT^2$ is a compact two-torus with Pontryagin dual $\hGamma=\ZZ^2$ (it is not difficult to imagine how one ought to proceed in other cases). This now means that there are in fact not so many choices for $\sigma$. Indeed, any two-cocycle on $\ZZ^2$ is cohomologous to a bicharacter of the form $\sigma((k,l),(m,n)):=\exp(\ii\theta(kn-lm))$ for some real number $\theta$ ({\em cf}. \cite{klep}).
}
\end{rem}

The action of $\Gamma$ on $\C^4$  yields in particular an action 
\begin{equation}\label{beta}
\beta:\Gamma\to\textup{Aut}(\A[\C^4])
\end{equation}
by $*$-algebra automorphisms.
Without loss of generality we may as well assume \cite{lvs:pfns} that it acts upon the generators $z_j$, $z_l^*$ according to
\begin{equation}\label{exp-act}
\beta:\Gamma\to\textup{Aut}(\A[\C^4]),\qquad\beta_r(z_1,z_2,z_3,z_4)=(e^{2\ii\pi r_1}z_1,e^{-2\ii\pi r_1}z_2,e^{2\ii\pi r_2}z_3,e^{-2\ii\pi r_2}z_4),
\end{equation}
for each element $r=(e^{2\ii\pi r_1},e^{2\ii\pi r_2})\in\Gamma$. Immediately we obtain a decomposition of the algebra $\A[\C^4]$ into spectral subspaces labelled by elements of $\hGamma$,
\begin{equation}\label{alg-sum}
\A[\C^4]=\bigoplus_{\zeta\in\hGamma}\, \A_\zeta[\C^4],
\end{equation}
where the $\zeta$-spectral subspace is defined to be
$$
\A_\zeta[\C^4]:=\left\{a\in\A[\C^4]~|~\beta_r(a)=\zeta(r)a~\text{for all}~r\in\Gamma\right\}.
$$
In contrast to the completed summation in eq.~\eqref{specde}, the direct sum \eqref{alg-sum} is purely algebraic.

Let us define a collection $\{V_\zeta~|~\zeta\in\hGamma\}$ of unitary multipliers by
\begin{equation}\label{units}
V_\zeta:=\lambda(\sigma^1_\zeta)^*\in \M(C^*(\Gamma))\subseteq \M(C_0(\C^4)\lcross\, \Gamma).
\end{equation}
Since the elements of the commutative algebra $\A[\C^4]$ are affiliated to the $C^*$-algebra $C_0(\C^4)$, it therefore makes sense to conjugate them by the multipliers $V_\zeta$, resulting in a new set of elements affiliated to $C_0(\C^4)\lcross\,\Gamma$. Given $\zeta\in\hGamma$ and $a\in \A_\zeta[\C^4]$ an element of the $\zeta$-spectral subspace, we define
$$
\hat a:=V_\zeta \,a\, V_\zeta^*.
$$
It is not difficult to see that the elements $\hat a$ are normal operators. We already know that they are affiliated to the crossed product $C_0(\C^4)\lcross\, \Gamma$ and so our task is to prove that they are also affiliated to the Landstad algebra $C_0(\C^4_\sigma)\subseteq \M(C_0(\C^4)\lcross\, \Gamma)$ for the $\sigma$-twisted dual action.

\begin{prop}
Let $f\in C_0(\C)$. Then for each $a\in \A_\zeta[\C^4]$ the element $f(\hat a)$ belongs to the multiplier algebra $\M(C_0(\C^4_\sigma))$.
\end{prop}

\proof First we check that $f(\hat a)$ is invariant under the twisted action $\hat\rho^\sigma$. In fact it is enough to check that $\hat a$ is invariant: we compute that
\begin{equation}\label{rsig}
\hat\rho_\xi^\sigma(\hat a)=\hat\rho_\xi^\sigma(V_\zeta)\,\hat\rho_\xi^\sigma(a)\,\hat\rho_\xi^\sigma(V_\zeta)^*.
\end{equation}
Individually computing each of the factors on the right hand side gives
\begin{align*}
\hat\rho_\xi^\sigma(a)&=\lambda(\sigma^1_\xi)\,\hat\rho_\xi(a)\,\lambda(\sigma^1_\xi)^*
=\lambda(\sigma^1_\xi)\,a\,\lambda(\sigma^1_\xi)^*,\\
\hat\rho_\xi^\sigma(V_\zeta)&=\hat\rho_\xi\left(\lambda(\sigma^1_\zeta)^*\right)
=\lambda(\sigma^1_{\zeta+\xi})^*
=\lambda(\sigma^1_{\zeta})^*\lambda(\sigma^1_{\xi})^*=V_\zeta\,\lambda(\sigma^1_{\xi})^*,
\end{align*}
where the first calculuation uses the fact that the element $a_\zeta$ is $\hat\rho$-invariant and the second uses the fact that $\lambda$ is a group homomorphism. Substituting these expressions into eq.~\eqref{rsig}, we deduce that
\begin{equation*}
\hat\rho_\xi^\sigma(\hat a)=V_\zeta\,\lambda(\sigma^1_\xi)^*\left(\lambda(\sigma^1_\xi)\,a\,\lambda(\sigma^1_\xi)^*\right)\lambda(\sigma^1_\xi)\,V_\zeta^*
=V_\zeta\,a \,V^*_\zeta=\hat a,
\end{equation*}
as required. Next we show that the map
\begin{equation}\label{n-con}
\Gamma\to M(C_0(\C^4)\lcross\,\Gamma),\qquad r\mapsto \lambda_r f(\hat a)\lambda_r^*
\end{equation}
is norm continuous. Indeed, we have that
$$
\lambda_r f(\hat a)\lambda_r^*=V_\zeta\lambda_rf(a)\lambda_r^* V_\zeta^*=V_\zeta f(\lambda_r a \lambda_r^*)V_\zeta^*
$$
and, since the function $f$ is continuous and vanishes at infinity, the map \eqref{n-con} is indeed norm continuous. This shows that the element $f(\hat a)$ satisfies the first two of Landstad's conditions in Def.~\ref{de:l-cons}, whence it is an  element of the multiplier algebra $\M(C_0(\C^4_\sigma))$.\endproof

\begin{prop}\label{pr:dens}
For each $a\in \A_\zeta[\C^4]$, the set 
$$
\mathcal{I}:=\left\{ f(\hat a) A^\sigma~|~f\in C_0(\C)\right\}
$$ 
is linearly dense in $A^\sigma$.
\end{prop}

\proof It is straightforward to check that $\mathcal{I}$ is invariant under the action $\hat\rho^{\sigma}$, since the latter is implemented by the unitary elements $\lambda_r$ for each $r\in \Gamma$. Let $g\in C_0(\C)$ be the function $g(z):=(1+z^*z)^{-1}$. Then $g(\hat a)=V_\zeta(1+a^*a)^{-1}V_\zeta^*$ and so we find that
\begin{align}\nonumber 
\left[ C^*(\Gamma) g(\hat a) A^\sigma C^*(\Gamma)\right]  &=\left[ C^*(\Gamma)V_\zeta(1+a^*a)^{-1}V_\zeta^*A^\sigma C^*(\Gamma)\right] \\
\label{den}&\subseteq \left[ C^*(\Gamma)\mathcal{I}C^*(\Gamma)\right],
\end{align}
where we have used the equality $C^*(\Gamma)V_\zeta=C^*(\Gamma)$. It is clear that the set $V_\zeta^* A^\sigma C^*(\Gamma)$ is linearly dense in $C_0(\C^4)\lcross\,\Gamma$ and so, since $a$ is affiliated to $C_0(\C^4)\lcross\,\Gamma$, it follows that the set 
$$
C^*(\Gamma)(1+a^*a)^{-1}V_\zeta^*A^\sigma C^*(\Gamma)
$$ 
is linearly dense in $C_0(\C^4)\lcross\,\Gamma$. Using the inclusion \eqref{den}, we deduce that the set $C^*(\Gamma)\mathcal{I}C^*(\Gamma)$ is linearly dense in $C_0(\C^4)\lcross\,\Gamma$. Using \cite[Lem.~2.6]{kasp} we obtain the linear density of $\mathcal{I}$ in $A^\sigma$.\endproof

Introducing the homomorphism of $C^*$-algebras
$$
\pi_a:C_0(\C)\to \M(C_0(\C^4_\sigma)),\qquad \pi_a(f):=f(\hat a),
$$
for each $j=1,\ldots,4$, we arrive at the desired theorem.

\begin{thm}
For each $a\in \A_\zeta[\C^4]$ the homomorphism $\pi_a$ is a morphism of $C^*$-algebras and the element $\hat a$ is a normal operator affiliated to $C_0(\C^4_\sigma)$.
\end{thm}

\proof By Prop.~\ref{pr:dens}, the norm closure of the set $\pi_a(C_0(\C))A^{\sigma}$ is equal to $A^\sigma$, which shows that $\pi_a:C_0(\C)\to A^\sigma$ is a morphism of $C^*$-algebras. Let $\iota \in C_0(\C)^{\eta}$ be the identity function defined by $\iota(z)=z$ for all $z\in\C$. Then applying the morphism $\pi_a$ to the function $\iota$, we find that $\pi_a(\iota)=\iota(\hat a)=\hat a\in (A^\sigma)^\eta$, whence the result.\endproof

Given an element of the coordinate algebra $a\in \A[\C^4]$, we therefore obtain a new element $\hat a$ by first decomposing $a$ into its spectral components according to eq.\eqref{alg-sum} and then taking the appropriate conjugate of each component. The latter theorem shows that this new element $\hat a$ is affiliated to the noncommutative $C^*$-algebra $C_0(\C^4_\sigma)$.

\begin{defn}\label{de:nc coord}
The coordinate algebra $\A[\C^4_\sigma]$ of noncommutative twistor space $\C^4_\sigma$ is the unital $*$-algebra 
$$
\A[\C^4_\sigma]:=\left\{ \hat a~|~a\in\A[\C^4]\right\}.
$$
\end{defn}

One finds that the commutation relations in the algebra $\A[\C^4_\sigma]$ are given by 
$$
\hat a_\zeta \hat a_\xi=\sigma(\zeta,\xi)^2\hat a_\xi \hat a_\zeta
$$
for each pair of elements $\hat a_\zeta\in \A_\zeta[\C^4_\sigma]$ and $\hat a_\xi\in \A_\xi[\C^4_\sigma]$ with respect to the spectral decomposition \eqref{alg-sum}. The affiliation relation between the $C^*$-algebra $C_0(\C^4_\sigma)$ (describing the topology of noncommutative twistor space) and the coordinate algebra $\A[\C^4_\sigma]$ (describing its algebraic structure) will be of particular importance in the final section, when we come to discuss the geometry of the noncommutative twistor fibration and the construction of instantons.

\section{Gauge Theory on Noncommutative Spin Manifolds}\label{se:gauge theory}

It is the founding principle of noncommutative geometry that the structure of a classical Riemannian spin manifold is encoded very succinctly in terms of an associated {\em spectral triple}, a notion which gives a very natural way to generalize many of the standard tools of differential geometry to the context of noncommutative spaces. In this section we recall the basic theory of spectral triples and examine the extent to which their symmetries describe gauge theories.

\subsection{Spectral triples and noncommutative spin geometry}\label{se:st} Let $A$ be a unital $C^*$-algebra.  According to the standard parlance of noncommutative geometry, we think of $A$ as the algebra of continuous functions on some underlying `virtual' compact topological space. The essence of a spin geometry on this noncommutative space is encapsulated by the following definition \cite{ac:book,mes}.

\begin{defn}\label{de:spec}
 A {\em spectral triple} $(\A,\mathcal{H},D)$ over a unital $C^*$-algebra $A$ consists of: 
\begin{enumerate}[\hspace{0.5cm}(i)]
\item a Hilbert space $\h$ equipped with a faithful representation $\mu_A:A\to \B(\h)$;
\item an unbounded self-adjoint operator $D:\Dom(D)\to\mathcal{H}$ with compact resolvent
\end{enumerate}
such that the {\em Lipschitz algebra}
$$
\A:=\{a\in A~|~[D,\mu_A(a)]\in\B(\h)\}
$$
is a dense $*$-subalgebra of $A$. Such a triple is called {\em even} if it is graded, i.e. if there exists a self-adjoint
operator $\Gamma:\mathcal{H}\to\mathcal{H}$ with $\Gamma^2=\id_\h$ such that $\Gamma D+D\Gamma=0$ and $\Gamma
\mu_A(a)=\mu_A(a)\Gamma$ for all $a\in A$. Otherwise the triple is said
to be {\em odd}.
\end{defn}

\begin{example}
Let $X$ be a compact Riemannian spin manifold and let $A=C(X)$ be the $C^*$-algebra of continuous complex-valued functions on $X$. Let $\h=L^2(M,\mathcal{S})$ be the Hilbert space of square-integrable sections of the spinor bundle $\mathcal{S}$, upon which $A$ acts faithfully by pointwise multiplication, and let $D:\Dom(D)\to\h$ denote the Dirac operator on $X$ determined by the Riemannian metric. The datum $(\A,\h,D)$ is called the {\em canonical spectral triple} over $A$, for which the Lipschitz algebra $\A$ is nothing other than the $*$-algebra of Lipschitz functions on $X$. The canonical spectral triple is even if and only if the underlying manifold $X$ is even-dimensional.
\end{example}

Next we come to the differential structure of a noncommutative spin manifold. Given a spectral triple $(\A,\mathcal{H},D)$ over a $C^*$-algebra $A$, one associates to it a
canonical first order differential calculus $(\Omega^1_DA,\D)$ in the following way.

\begin{defn}
The $\A$-$\A$-bimodule
$\Omega^1_DA$ of differential one-forms over $A$ is defined to be
\begin{equation}\label{commrep}
\Omega^1_DA:=\{\omega=\sum_j a^j_0[D,a^j_1]~|~a^j_0, a^j_1\in \A\},
\end{equation}
where the summations defining the one-forms $\omega\in\Omega^1_DA$ are taken to be convergent in the norm topology on $\B(\h)$. The exterior derivative $\D:\A\to\Omega^1_DA$ is defined by the formula $\D a:=[D,a]$ for each $a\in \A$.
\end{defn}

The exterior derivative $\D:\A\to\Omega^1_DA$ extends in the usual way \cite{ac:book} by imposing a graded Leibniz rule and requiring that $\D^2=0$, followed by making a quotient by Connes' ideal of `junk' differential forms, yielding a complex of higher differential forms $$\Omega^*_D A=\bigoplus_k\,\Omega^k_DA$$ for $k=0,1,2,\ldots$, whose differential we denote by $\D:\Omega^k_DA\to\Omega^{k+1}_DA$.

\begin{example}
In the case of the canonical spectral triple over a compact spin manifold $X$, this definition of $\Omega^1_DA$ recovers the $\A$-bimodule $\Omega^1_D(X)$ of continuous one-forms on $X$, which extends to a complex of higher differential forms in the standard way, of course taking into account the fact that the exterior derivative is an unbounded operator and hence only has dense domain.
\end{example}

Combining the canonical spectral triple over a  Riemannian spin manifold $X$ with the Landstad-Kasprzak deformation theory gives a natural way of obtaining noncommutative spin geometries. Indeed, let $X$ be a compact Riemannian spin manifold and let $(\A,\H,D)$ denote the canonical spectral triple over $A=C(X)$. Suppose that $X$ is equipped with an isometric action of a compact Abelian group $\Gamma$ and denote by  $\alpha:\Gamma\to \textup{Aut}(A)$ the corresponding action of $\Gamma$ on $A$. Let $\sigma:\hGamma\times\hGamma\to\TT$ be a two-cocycle on $\hGamma$ and write $(A^\sigma,\Gamma,\alpha^\sigma)$ for the twisted covariant system obtained from the classical covariant system $(A,\Gamma,\alpha)$.

The passage from $A$ to $A^\sigma$ deforms the topological space $X$ into a virtual noncommutative space $X_\sigma$, in the sense that we regard $A^\sigma=C(X_\sigma)$ as the $C^*$-algebra of continuous functions on $X_\sigma$. However, we otherwise leave the spin structure alone, 
i.e. we leave unchanged the Hilbert space and the Dirac operator of the canonical spectral triple. 
The isometric action of $\Gamma$ on $X$ lifts to the spinor bundle $\mathcal{S}$ (although not canonically so, there are in general many inequivalent choices, {\em cf}. for example \cite{dab:spin}).  Given such a lift, let us write $\mu_\Gamma:\Gamma\to \mathbb{B}(\H)$ for the corresponding unitary representation on the Hilbert space of square-integrable sections and assume that the action is smooth, in the sense that $\mu_\Gamma(\Dom(D))\subseteq\Dom(D)$. Then we have
\begin{equation}\label{covs}
 \mu_\Gamma(g)\,\mu_A(a) \,\mu_\Gamma(g^{-1})=\mu_A(\alpha_g(a)),\qquad \mu_\Gamma(g)\,D \,\mu_\Gamma(g^{-1})=D, 
\end{equation}
for all $g\in \Gamma$. The first condition in \eqref{covs} says that we have a covariant representation of $(A,\Gamma,\alpha)$ on the Hilbert space $\H$ and hence a representation of the crossed product $A\lcross_\alpha\,\Gamma$. Using the isomorphism $A\lcross_\alpha\,\Gamma\simeq A^\sigma\lcross_{\alpha^\sigma}\,\Gamma$, we obtain a representation of the twisted covariant system $(A^\sigma,\Gamma,\alpha^\sigma)$ on the same Hilbert space $\H$. 

The second condition in \eqref{covs} implies that, by leaving the Hilbert space and the Dirac operator unchanged, we obtain a spectral triple $(\A^\sigma,\H,D)$ over the deformed $C^*$-algebra $A^\sigma=C(X_\sigma)$ and hence a noncommutative spin geometry on the quantum space $X_\sigma$. Here we write $\A^\sigma$ for the Lipschitz algebra of $A^\sigma$ in the sense of Def.~\ref{de:spec} (which should not be confused with the noncommutative coordinate algebras obtained as in Def.~\ref{de:nc coord}). The details are checked just as in \cite{cl:id,blvs}. We write $\Omega^*_D(X_\sigma)$ for the differential calculus determined by the Dirac operator. Since the Dirac operator is unchanged, we call $(\A^\sigma,\H,D)$ the spectral triple over $A^\sigma$ obtained by {\em isospectral deformation} of the classical spectral triple $(\A,\H,D)$ over $A$.

\subsection{Gauge theory of noncommutative spin manifolds} In classical Riemannian geometry, a unitary gauge theory over a given manifold $X$ consists of a Hermitian vector bundle over $X$ equipped with a metric-compatible connection. Here we recall how to generalize this idea to noncommutative spin manifolds.

Let $A$ be a unital $C^*$-algebra with norm $\|\cdot\|_A$. Recall that the notion of a Hermitian vector bundle over the underlying noncommutative space is given by that of a Hilbert $A$-module.

\begin{defn}
A right Hilbert $A$-module, denoted $\mathpzc{E}\leftrightharpoons A$, is a right $A$-module $\mathpzc{E}$ equipped with a right $A$-valued inner product 
\begin{equation}\label{hilb-ip}
\la\cdot,\cdot\ra_A:\mathpzc{E}\times\mathpzc{E}\to A
\end{equation}
such that $\mathpzc{E}$ is complete in the norm $\|e\|^2=\|\la e,e\ra\|_A$, where $e\in\mathpzc{E}$.
\end{defn}

The module $\mathpzc{E}$ plays the role of the space of continuous sections of an underlying vector bundle. Indeed, given a compact Hausdorff space $X$, every (finitely generated and full) Hilbert module $\mathpzc{E}\leftrightharpoons C(X)$ is isomorphic to the right $C(X)$-module $\Gamma(X,E)$ of continuous sections of some Hermitian vector bundle $E\to X$. We write $\End^*_A(\mathpzc{E})$ for the $C^*$-algebra of adjointable linear operators \cite{Lance} on a right Hilbert module $\mathpzc{E}\leftrightharpoons A$. If $\mathpzc{E}$ is equipped with a representation $B\to \End^*_A(\mathpzc{E})$, we write $B\to\mathpzc{E}\leftrightharpoons A$ and say that $\mathpzc{E}$ is a {\em Hilbert $B$-$A$ bimodule}.




\begin{rem}\textup{
Just as we did for spectral triples in Def.~\ref{de:spec}, we allow Hilbert modules $\mathpzc{E}\leftrightharpoons A$ and Lipschitz modules $\E\leftrightharpoons \A$ to be $\ZZ_2$-graded, which in turn imposes a grading on the $C^*$-algebra $\End^*_A(\mathpzc{E})$. 
In doing so, when taking tensor products of such modules we shall always mean the graded tensor product; similarly, the tensor product of graded linear operators will always be the graded one \cite{higroe}.
}
\end{rem}


Let $(\A,\H,D)$ be a spectral triple over a unital $C^*$-algebra $A$. The following definition gives us an appropriate notion of connection on a noncommutative vector bundle.

\begin{defn}
A {\em Hermitian connection} $\n$ on a right Hilbert module $\mathpzc{E}\leftrightharpoons A$ is an unbounded linear map $\n:\mathfrak{Dom}(\n)\to \mathpzc{E}\otimes_{A}\Omega^1_DA$ with dense domain $\Dom(\n)\subseteq \mathpzc{E}$, obeying the conditions
\begin{align*}
\n(e a)&=(\n e)a+e\otimes \D a, \\
\D \la e,f\ra_A&=\la\n e,f\ra_A+\la e,\n f\ra_A,
\end{align*}
for all $e,f\in \mathfrak{Dom}(\n)$ and all $a\in \A$.
\end{defn}


Now that we have a suitable notion of a vector bundle with connection, we are ready to discuss its gauge theory \cite{ac:book,ac:fncg,bmvs}. Let $(\A,\H,D)$ be a spectral triple over a unital $C^*$-algebra $A$ and let $\mathpzc{E}\leftrightharpoons A$ be a finitely generated right Hilbert $A$-module. In this paper we shall consider two types of gauge transformations which arise naturally in noncommutative geometry. The first is the obvious generalization of the unitary gauge group of a classical Hermitian vector bundle.

\begin{defn}
The {\em external gauge group} of the module $\mathpzc{E}\leftrightharpoons A$ is the group
$$
\mathcal{G}_e(\mathpzc{E}) := \{ U \in \End^*_{A}(\mathpzc{E}) ~|~ UU^*=\id_{\mathpzc{E}} =U^*U\}
$$
of unitary endomorphisms of the module $\mathpzc{E}\leftrightharpoons A$.
\end{defn}

The external gauge group $\mathcal{G}_e(\mathpzc{E})$ of the module $\mathpzc{E}\leftrightharpoons A$ acts upon a Hermitian connection $\n:\Dom(\n)\to \mathpzc{E}\otimes_A\Omega^1_{D} A$ in the usual way by conjugation, 
\begin{equation}\label{gauge-act}
\n\mapsto\n^U:= (U\otimes \id_{\Omega^1})\n U^*,
\end{equation} 
for each $U\in\mathcal{G}_e(\mathpzc{E})$, yielding a new connection $\n^U:\Dom(\n^U)\to \mathpzc{E}\otimes_A\Omega^1_{D} A$.

The second type of gauge transformation we shall need arises from the notion of Morita equivalence between $C^*$-algebras. With $(\A,\H,D)$ as above, recall that any $C^*$-algebra $B$ which is Morita equivalent to $A$ is necessarily isomorphic to the algebra of adjointable endomorphisms of some finitely generated right Hilbert $A$-module, that is to say
$$
B=\End^*_A(\mathpzc{E})
$$
for some $\mathpzc{E}\leftrightharpoons A$. Given a choice $\n: \Dom(\n)\to \mathpzc{E}\otimes_A\Omega^1_{D}A$ of Hermitian connection, one may construct a new spectral triple $(\cB, \H_\mathpzc{E},D_\n)$ over the $C^*$-algebra $B$ by setting
\begin{equation}\label{stb}
\H_\mathpzc{E}:=\mathpzc{E}\otimes_A\H,\qquad D_\n:=\id\otimes D +\n\otimes \id,
\end{equation}
with $\cB$ the corresponding Lipschitz algebra. This construction first appeared in \cite{ac:fncg}. 

In the special case where $B=\mathpzc{E}=A$, a connection is given by an unbounded linear map
$$
\n:\A\to\Omega^1_{D} A
$$
and we automatically have that $\n=\D+\omega$ for some self-adjoint one-form $\omega=\omega^*\in\Omega^1_{D}A$. Using the identification $\H:=\mathpzc{E}\otimes_A\H$ we have that $D_\n=D+\omega$.  In this special case, we write $D_\omega:=D_\n$, so that a choice of connection results in a perturbation $D\mapsto D_\omega$ of the Dirac operator on $A$.

\begin{defn}
The {\em internal gauge group} of the spectral triple $(\A,\H,D_\omega)$ is the group 
$$
\textup{U}(A):=\{u\in A~|~u^*u=1_A=uu^*\}
$$
of unitary elements of the $C^*$-algebra $A$.
\end{defn}

The internal gauge group $\textup{U}(A)$ acts upon the spectral triple $(\A,\H,D_\omega)$ by unitary equivalences, according to the replacement
$$
D_\omega\mapsto D_\omega^u:=uD_\omega u^*
$$
or, equivalently, according to the familiar transformation rule
\begin{equation}\label{int-act}
\omega\mapsto \omega^u:=u\omega u^*+ u[D,u^*]
\end{equation}
for each $u\in \textup{U}(A)$. When the algebra $A$ is commutative, the one-forms commute with functions and so the fluctuations $D\mapsto D_\omega$ and the action \eqref{int-act} of the internal gauge group $\textup{U}(A)$ are trivial \cite{ac:fncg,bl:adhm}. The internal gauge group is therefore not visible in gauge theory over a commutative algebra: it is a purely `quantum' phenomenon.

\begin{rem}
\textup{
Although the external gauge group $\mathcal{G}_e(\mathpzc{E})$ and the internal gauge group $\textup{U}(A)$ may at first appear to be very different and irreconcilable entities, it turns out that there is in fact a very natural unifying framework in which these two notions of gauge transformation may be brought together \cite{bmvs}. More on this will be reported elsewhere.
}
\end{rem}

\section{The Noncommutative Topology of Instanton Gauge Fields}\label{se:final}

This final section is dedicated to the study of instantons on the noncommutative four-sphere $S^4_\sigma$ constructed earlier in the paper. We begin with a brief review of the basic definitions of anti-self-dual fields (instantons) and their gauge theory. We then recall the construction of instantons on the classical four-sphere $S^4$, to which we apply the  Landstad-Kasprzak deformation functor.

\subsection{Instantons on noncommutative four-spheres} 
Following Rem.~\ref{re:tt}, we recall that the topological deformation of $S^4$ described in \S\ref{se:def-tw} was implemented by the action of a  two-torus $\Gamma=\TT^2$ acting continuously on the four-sphere $S^4$. However, in order to preserve the spin geometry of $S^4$ and obtain an isospectral noncommutative manifold as in \S\ref{se:st}, we assume in addition that $\Gamma$ acts upon $S^4$ by Euclidean isometries. The very form of the isometry group of $S^4$ justifies our assumption that $\Gamma$ is a torus of rank two and not of higher rank.

In this section we shall simplify our notation and write $A:=C(S^4_\sigma)$ for the Landstad--Kasprzak deformation of the classical $C^*$-algebra $C(S^4)$, writing $(\A,\H,D)$ for the spectral triple over $A$ obtained by isospectral deformation: this determines the noncommutative Riemannian spin structure of the quantum sphere $S^4_\sigma$. The corresponding Hodge structure is obtained just as in \cite{blvs}. Let us recall the definition.

\begin{defn}\label{def:hodge}
The Hodge $*$-operator on $S^4_\sigma$ is the linear map $*:\Om^r(S^4_\sigma)\to\Om^{4-r}(S^4_\sigma)$ defined for each $\omega_1, \omega_2\in\Om^r(S^4_\sigma)$ by the formula
$$
\omega_1\wedge(*\omega_2)=(\omega_1,\omega_2)\varpi,
$$
where $(\,,\,)$ is the Hermitian structure on differential forms and $\varpi$ denotes the volume form.
\end{defn}

As an operator $*:\Om^r(S^4_\sigma)\to\Om^{4-r}(S^4_\sigma)$, the Hodge $*$-operator obeys $*^2=(-1)^r\id$ and so, in particular, it maps the space $\Om^2(S^4_\sigma)$ of two-forms onto itself. One therefore has a direct sum decomposition into eigenspaces
\begin{equation}\label{eig-decomp}
\Om^2(S^4_\sigma)=\Om^2_+\oplus \Om^2_-,
\end{equation}
where $\Om^2_\pm:=\{\omega\in\Om^2(S^4_\sigma)~|~*\omega=\pm\omega\}$. The elements of $\Om^2_+$ are said to be {\em self-dual}; the elements of $\Om^2_-$ are said to be {\em anti-self-dual}.

Now let $\mathpzc{E}\leftrightharpoons A$ be a finitely generated right Hilbert $A$-module, which we assume arises as a deformation of the space of sections of a $\Gamma$-equivariant vector bundle over $S^4$ in the sense of \cite{blvs}. Let $\n:\Dom(\n)\to\mathpzc{E}\otimes_{A}\Om^1(S^4_\sigma)$ be a Hermitian connection on $\mathpzc{E}$ and consider the spectral triple $(\cB,\H_B,D_B)$ over the $C^*$-algebra $B=\End^*_A(\mathpzc{E})$ described in eq.~\eqref{stb}.

The assumption that the module $\mathpzc{E}$ is $\Gamma$-equivariant implies that it is in fact an $A$-bimodule and in particular that there is an isomorphism $\textup{Hom}_{A}(\mathpzc{E},\mathpzc{E}\otimes_A\Om^2(S^4_\sigma))\cong\End^*_{A}(\mathpzc{E})\otimes_A\Om^2(S^4_\sigma)$ ({\em cf.} \cite{blvs} for a careful explanation of these facts, in particular that $\End^*_{A}(\mathpzc{E})$ is also an $A$-$A$-bimodule and so the right hand side of the latter identification is well defined). As a consequence, the Hodge $*$-operator extends to a linear map
$$
\id\otimes *:\End^*_{A}(\mathpzc{E})\otimes_A\Om^2(S^4_\sigma)\to \End^*_{A}(\mathpzc{E})\otimes_A\Om^2(S^4_\sigma)
$$
on endomorphism-valued two-forms.

\begin{defn}
A Hermitian connection $\n:\Dom(\n)\to\mathpzc{E}\otimes_{A}\Om^1(S^4_\sigma)$ on $\mathpzc{E}$ is said to be an {\em instanton} if its curvature $F_\n\in\End^*_{A}(\mathpzc{E})\otimes_A\Om^2(S^4_\sigma)$ is an anti-self-dual two-form.
\end{defn}


Under the action \eqref{gauge-act} of the external gauge group $\mathcal{G}_e(\mathpzc{E})$, the curvature $F_{\n}$ of the connection $\n$ transforms according to the rule
$$
F_{\n}\mapsto F_{\n{}^U}=(U\otimes\id_{\Omega^2})F_{\n} (U^*\otimes \id_{\Omega^2}),\qquad U\in\mathcal{G}_e(\mathpzc{E}),
$$
from which it follows that, if $\n$ is an instanton, then so is the gauge-transformed connection $\n^U$ (due to the fact that, in the tensor product $\End^*_{A}(\mathpzc{E})\otimes_A\Om^2(S^4_\sigma)$, the endomorphism $U$ acts upon the factor $\End^*_{A}(\mathpzc{E})$, whereas the Hodge operator acts only upon $\Om^2(S^4_\sigma)$, {\em cf}. \cite{blvs}).

The action of the two-torus $\Gamma$ upon the classical sphere $S^4$ is by Euclidean isometries and so it commutes with the Hodge $*$-operator $*:\Omega^2_D(S^4)\to\Omega^2_D(S^4)$, whence the same is true of the Hodge operator $*:\Omega^2_DA\to\Omega^2_DA$ on the noncommutative sphere $S^4_\sigma$.  In this case the internal gauge group $\U(A)$ is interpreted as being the group of continuous functions on $S^4_\sigma$ with values in the circle group $\U(1)$. 

The action \eqref{int-act} of the internal gauge group $\U(A)$ upon the spectral triple $(\A,\H,D)$, by unitary equivalences therefore preserves the  eigenspace decomposition \eqref{eig-decomp} of two-forms into self-dual and anti-self-dual components, since the same is true in the classical case. It follows that the internal gauge group also preserves the anti-self-dual curvature condition and so it has a well defined action upon the space of instanton connections.

\begin{rem}\textup{The philosophy that we shall adopt in the present paper is that, if we wish to construct the space of gauge equivalence classes of instanton connections on a given vector bundle over $S^4_\sigma$, we should divide the space of all such instanton connections not just by the group of external gauge transformations (as we would do in classical geometry) but also by the action of the internal gauge group ({\em cf}. \cite{bl:mod}).
}
\end{rem}

\subsection{Construction of instantons on the classical sphere}\label{se:cl-adhm} In order to construct instantons on the noncommutative sphere $S^4_\sigma$, our strategy will be to start with the construction of instantons on the classical sphere $S^4$ and then to see what happens to this construction under the deformation. Let  $\mathpzc{E}\leftrightharpoons A$ be a finitely generated right Hilbert module over the $C^*$-algebra $C(S^4)$, necessarily isomorphic to the space $\Gamma(S^4,E)$ of continuous sections of some Hermitian vector bundle $E\to S^4$.  

Let us further specify the topology of our vector bundle by assuming that $\mathpzc{E}\leftrightharpoons A$ has Chern numbers $\textup{ch}_0(\mathpzc{E})=n$ and $\textup{ch}_2(\mathpzc{E})=k$, the noncommutative analogues of the rank and the first and second Chern classes \cite{ac:book,gbvf} (the first Chern class $\textup{ch}_1(\mathpzc{E})$ is automatically zero, since it takes values in the trivial homology group $H^2(S^4,\mathbb{Z})$).

As already mentioned, in this case the action of the internal gauge group is trivial and so the gauge freedom for connections on $\mathpzc{E}$ is given by the external gauge group $\mathcal{G}_e(\mathpzc{E})$. Given a connection $\n$ we write $[\n]$ for its equivalence class under the action of this gauge group.

\begin{defn}
The {\em moduli space of instantons} on $\mathpzc{E}$ is defined to be the set
$$
\mathcal{M}^{k,n}:=\left\{  [\n]~|~*(F_\n)=-F_\n\right\}
$$
of gauge equivalence classes of connections $\n:\Dom(\n)\to\mathpzc{E}\otimes_{A}\Omega^1_D(S^4)$ with anti-self-dual curvature.
\end{defn}

The moduli space $\mathcal{M}^{k,n}$ has a natural topology inherited from that of the affine space of all connections on $\mathpzc{E}$ although, of course, it is far from being Hausdorff, due to the presence of singularities coming from the fixed points of the gauge group action.

As remarked in the introduction, the construction of instantons is equivalent to the construction of holomorphic vector bundles over $\CP^3$. The crucial ingredient in doing so is the following.

\begin{defn}\label{de:monad}
Let $k,n\in\ZZ$ be fixed positive integers. A {\em monad} with indices $k,n$ over homogeneous twistor space $\C^4$
is a complex of free right $C_0(\C^4)$-modules
\begin{equation}\label{monad}
M:\qquad 0\to H\otimes C_0(\C^4) \xrightarrow{\rho_z} K\otimes C_0(\C^4)
\xrightarrow{\tau_z} L\otimes C_0(\C^4)\to 0,
\end{equation}
where $H$, $K$ and $L$ are complex vector spaces of dimensions $k$,
$2k+n$ and $k$ respectively, such that the right module maps $\rho_z$, $\tau_z$
are linear in the coordinates $z_1,\ldots,z_4$ of $\C^4$.
\end{defn}

In the latter definition, we think of the coordinate functions $z_1,\ldots,z_4\in C(\C^4)$ as
unbounded operators affiliated to the $C^*$-algebra $C_0(\C^4)$. As such, they act upon $C_0(\C^4)$ via their $\mathfrak{z}$-transforms,
$$
z_j\in C(\C^4)~\mapsto ~\mathfrak{z}(z_j):=z_j(1+z_jz_j^*)^{-1/2}\in \M(C_0(\C^4)),
$$
for each $j=1,\ldots,4$. The module maps $\rho_z$, $\tau_z$ being linear in the coordinate functions $z_1,\ldots,z_4$ means that they have the form
\begin{equation}\label{mon-maps}
\rho_z=\sum_j M_j\otimes z_j,\qquad \tau_z=\sum_l N_l\otimes z_l,
\end{equation}
for complex-valued matrices $M_j\in \M_{k,2k+n}(\C)$ and $N_l\in\M_{2k+n,k}(\C)$ with $j,l=1,\ldots,4$. 
This gives meaning to the composition of the maps $\rho_z$ and $\tau_z$, {\em viz}.
$$
\tau_z\circ\rho_z=\sum_{j,l}N_j M_l\otimes z_jz_l.
$$
For $j\leq l$ the quantities $z_jz_l$ are all linearly independent, from which the condition $\tau_z\circ\rho_z=0$ becomes explicitly
\begin{equation}\label{adhm eqs}
\sum\nolimits_b\left(N_j^{cb}\, M_l^{bd}\,+\,N_l^{cb} \,M_j^{bd}\right)=0
\end{equation} 
for all $j,l=1,\ldots,4$ and all $c,d=1,\ldots,k$. In the next definition, we denote the vector space duals of $H$, $K$, $L$ by $H'$, $K'$, $L'$, respectively.

\begin{defn}
Given a monad \eqref{monad}, we define its {\em conjugate
monad} to be the complex of free right modules
\begin{equation}\label{con-mon}
M':\qquad 0\to L'\otimes C_0(\C^4)\xrightarrow{\tau_{\J(z)}'} K'\otimes
C_0(\C^4)\xrightarrow{\rho_{\J(z)}'} H'\otimes C_0(\C^4),
\end{equation}
where $\J$ is the quaternionic involution defined in eq.~\eqref{J}
and $\tau_{z}'$, $\sigma_{z}'$ are the adjoint maps
defined by the canonical Hermitian structures.
\end{defn}

As with all constructions in linear algebra, the explicit form of a monad depends upon the choice of basis in each of the vector spaces $H$, $K$ and $L$. The following definition gives us a precise formulation of when two monads are equivalent.

\begin{defn}\label{de:eq-mon}
Monads $M$, $\widetilde M$ are said to be {\em equivalent} if
there is a commutative diagram
$$
\begin{CD} 
M:\qquad @. H\otimes C_0(\C^4) @>\rho_z >> K\otimes C_0(\C^4) @>\tau_z >> L\otimes C_0(\C^4)
\\ @. @VV u\otimes\id V @VV v\otimes\id V @VV w\otimes\id   V\\  \widetilde{M}:\qquad @. \widetilde{H}\otimes C_0(\C^4) @>\tilde\rho_z >> \widetilde{K}\otimes C_0(\C^4) @>\tilde\tau_z >> \widetilde{L}\otimes C_0(\C^4)\end{CD}
$$
for some invertible linear transformations $u\in\GL(H,\widetilde{H})$, $v\in\GL(K,\widetilde{K})$ and $w\in\GL(L,\widetilde{L})$. We denote this equivalence relation by $\sim$.
\end{defn}

\begin{defn}\label{de:scmon}
A monad \eqref{monad} which is equivalent to its conjugate
\eqref{con-mon} is said to be {\em self-conjugate}. We
write ${\sfM}^{k,n}$ for the set of all self-conjugate monads with indices
$k,n\in \ZZ$.
\end{defn}

The self-conjugacy condition imposes a certain relationship between the matrices $M^j$ and $N^l$ for $j,l=1,\ldots,4$. Indeed, the adjoint maps $\rho_{\J(z)}'$ and $\tau_{\J(z)}'$ have the explicit form
$$
\rho_{\J(z)}'=\sum_j M_j{}^* \otimes \J(z_j)^*,\qquad \tau_{\J(z)}'=\sum_l N_l{}^* \otimes \J(z_l)^*,
$$ 
where $M_j{}^*$ and $N_l{}^*$ denote the conjugate transpose matrices. From this one easily finds that, for a monad to be self-conjugate, we must have
\begin{equation}\label{monad star}
N_1=-M_2{}^*, \quad N_2=M_1{}^*, \quad N_3=-M_4{}^*, \quad
N_4=M_3{}^*.
\end{equation}
In this way, we think of the matrix elements $M_j^{ab}$ and $N_l^{cd}$ (modulo the relations \eqref{adhm eqs} and \eqref{monad star}) as parametrizing the space ${\sfM}^{k,n}$ of self-conjugate monads.

Monads are the input data for the construction of instantons on $S^4$. Indeed, the ADHM construction \cite{adhm:ci} provides an algorithm for converting monads over $\C^4$ into vector bundles over $S^4$ equipped with an instanton connection. Given a monad \eqref{monad}, its cohomology is a holomorphic vector bundle over $\CP^3$ with rank $n$ and second Chern number $k$. The self-conjugacy condition (of being invariant under the quaternionic involution \eqref{J}) implies that this bundle is the pull-back of a Hermitian vector bundle over $S^4$ with the same rank and Chern number; the canonical connection on this bundle has anti-self-dual curvature and every such connection arises from a monad in this way.


It is known \cite{adhm:ci} that any two instanton connections are equivalent under the action of the gauge group $\mathcal{G}_e(\mathpzc{E})$ if and only if they arise from monads which are equivalent in the sense of Def.~\ref{de:eq-mon}. In this way, the moduli space of instantons on a given Hermitian vector bundle over $S^4$ is given by the quotient
$$
\mathcal{M}^{k,n}\simeq\mathsf{M}^{k,n}/\sim.
$$
As already mentioned, since the algebra $A=C(S^4)$ is commutative, the action \eqref{int-act} of the internal gauge group $\textup{U}(A)$ is trivial \cite{ac:fncg,bl:adhm} and so plays no role in the classical case.

The subtle but important feature of the construction is that it replaces the rather awkward infinite dimensional space of all instanton connections by the much more manageable finite dimenional space $\mathsf{M}^{k,n}$. Similarly, the infinite dimensional gauge group $\mathcal{G}_e(\mathpzc{E})$ is replaced by the finite dimensional group
\begin{equation}\label{fd-group}
G=\GL(H)\times \GL(K)\times \GL(L).
\end{equation}
Obtaining the moduli space in this way as the quotient of finite dimensional spaces is a far easier computation \cite{don}. Our next step is to investigate the effect of the Landstad-Kasprzak deformation theory of the twistor fibration upon the moduli space $\mathcal{M}^{k,n}$ and its topology.

\subsection{A noncommutative space of monads} We now look to modify this formulation of the space of monads over classical twistor space $\C^4$, with a view to obtaining an analogous picture of the space of monads over noncommutative twistor space $\C^4_\sigma$ and hence a construction of instantons on the quantum sphere $S^4_\sigma$. The first step is to deform each of the free modules appearing in the complex \eqref{monad} into free $C_0(\C^4_\sigma)$-modules.

In order to deform the algebra $C_0(\C^4)$, we used the action
\begin{equation}\label{rec-act}
\beta:\Gamma\to\textup{Aut}(C_0(\C^4)),\qquad (\beta_r(a))(x):=a(r^{-1}x),
\end{equation}
where $r\in\Gamma$, $a\in C_0(\C^4)$ and $x\in\C^4$. Given a finite-dimensional vector space $H$, the action \eqref{rec-act} extends to an action of $\Gamma$ on the free right module $H\otimes C_0(\C^4)$ by
\begin{equation}\label{mod act}
\beta^H:\Gamma\to\textup{End}^*_{C_0(\C^4)}(H\otimes C_0(\C^4)),\qquad \beta^H_r:=\id\otimes\beta_r,
\end{equation}
for each $r\in \Gamma$. Given a monad $M$ and an element $r\in \Gamma$, we obtain a new monad by acting in this way upon each of the modules in the complex \eqref{monad}, that is to say
$$
\begin{CD} 
M:\qquad @. H\otimes C_0(\C^4) @>\rho_z >> K\otimes C_0(\C^4) @>\tau_z >> L\otimes C_0(\C^4)
\\ @. @VV \beta^H_r V @VV \beta^K_r V @VV \beta^L_r   V\\  M^r:\qquad @.  H\otimes C_0(\C^4) @>\rho^r_z >> K\otimes C_0(\C^4) @>\tau^r_z >> L\otimes C_0(\C^4),
\end{CD}
$$
where the right module maps in the sequence $M^r$ are defined by
$$
\rho^r_z:=\beta^K_r\circ\rho_z\circ\beta^H_{r^{-1}},\qquad \tau^r_z:=\beta^L_r\circ\tau_z\circ\beta^K_{r^{-1}},
$$
for each $r\in \Gamma$. It is clear that $\tau^r_z\circ\rho^r_z=0$, so the group action preserves the monad condition. Moreover, since the action \eqref{rec-act} commutes with the quaternionic structure \eqref{J}, if the given monad $M$ is self-conjugate, then so is the resulting monad $M^r$, giving us a natural action of $\Gamma$ on the space $\mathsf{M}^{k,n}$. Let us examine this action in more detail: to do so we introduce the notation
$$
\mu^r=(\mu^r_j):=(e^{2\ii\pi r_1},e^{-2\ii\pi r_1},e^{2\ii\pi r_2},e^{-2\ii\pi r_2})
$$
so that, in the notation of eq.~\eqref{exp-act}, we have $\beta_r(z_j)=\mu^r_jz_j$ for each $j=1,\ldots,4$.

\begin{lem}\label{le:c-map}
For each $r\in \Gamma$ we have 
\begin{equation}\label{g-mon}
\rho^r_z=\sum_j \mu^r_j M_j\otimes z_j,\qquad \tau^r_z=\sum_l \mu^r_l N_l\otimes z_l.
\end{equation}
\end{lem}

\proof For each $r\in\Gamma$ and all $v\in H$, $a\in  C_0(\C^4)$,
it is straightforward to compute that
\begin{align*}
\left(\beta^K_r\circ\rho_z\circ\beta^H_{r^{-1}}\right)(v\otimes a)&=\beta^K_r\left(\sum_j M_j v\otimes \mathfrak{z}(z_j)\beta_{r^{-1}}(a)\right) \\
&=\sum_j M_j v\otimes \beta_r(\mathfrak{z}(z_j))a \\
&=\sum_j \mu^r_jM_j v\otimes \mathfrak{z}(z_j)a.
\end{align*}
A similar computation establishes the analogous result for the map $\tau_z$.\endproof

Acting upon the modules according to \eqref{mod act} therefore induces a continuous action of $\Gamma$ on the space $\mathsf{M}^{k,n}$, defined by eq.~\eqref{g-mon}. In particular this means that, if we want to deform each of the modules in the complex \eqref{monad} into $C_0(\C^4_\sigma)$-modules, we cannot ignore the fact that the $\Gamma$-action \eqref{mod act} also acts on the maps {\em between} these modules. 

As a submanifold of $\C^{4k(2k+n)}$, the space $\mathsf{M}^{k,n}$ of self-conjugate monads over $\C^4$ is a locally compact and Hausdorff topological space. We shall therefore study it (and its deformations) from the point of view of the $C^*$-algebra $C_0({\sfM}^{k,n})$ of continuous functions vanishing at infinity. The action \eqref{g-mon}  immediately gives rise by pull-back to an action 
\begin{equation}\label{monact}
\tilde\beta:\Gamma\to \textup{Aut}(C_0(\mathsf{M}^{k,n}))
\end{equation}
of $\Gamma$ by automorphisms of the $C^*$-algebra $C_0(\mathsf{M}^{k,n})$. In applying the deformation procedure to twistor space, it follows that the space of monads is automatically deformed. With the same cocycle $\sigma:\hGamma\times\hGamma\to\mathbb{T}$ that we used in \S\ref{se:def-tw}, we arrive at the following definition.

\begin{defn}
We write $C_0(\mathsf{M}^{k,n}_\sigma)$ for the Landstad-Kasprzak deformation of the $C^*$-algebra $C_0(\mathsf{M}^{k,n})$ along the $\Gamma$-action \eqref{monact} via the two-cocycle $\sigma$. We think of $C_0(\mathsf{M}^{k,n}_\sigma)$ as the algebra of continuous functions vanishing at infinity on the noncommutative space $\mathsf{M}^{k,n}_\sigma$.
\end{defn}

Now we turn to the algebraic structure of the space of monads. Just as in \cite{bl:mod}, we regard the matrix elements $M_j^{ab}$ and $N_l^{cd}$ 
as coordinate functions on the space ${\sfM}^{k,n}$. We write $\A[{\sfM}^{k,n}]$ for the coordinate algebra of ${\sfM}^{k,n}$, that is to say, the commutative unital $*$-algebra generated by the matrix elements $M_j^{ab}$ and their conjugates $M_j^{ab}{}^*$, subject to the relations \eqref{adhm eqs}. In these terms, the set of all pairs of module maps $\rho_z$, $\tau_z$ appearing in eq.~\eqref{monad} may be written
\begin{align}\label{module maps}
\rho_z:H\otimes C_0(\C^4)\to \A[{\sf M}^{k,n}] \otimes K\otimes C_0(\C^4),\qquad \rho_z&: v\otimes a \mapsto \sum\nolimits_{j}
M_j \otimes v \otimes \mathfrak{z}(z_j) a, \\
\label{modmaps2}\tau_z:K\otimes C_0(\C^4)\to \A[{\sf M}^{k,n}] \otimes L\otimes C_0(\C^4), \qquad \tau_z&: w\otimes a \mapsto \sum\nolimits_{j}
N_j \otimes w \otimes \mathfrak{z}(z_j) a, 
\end{align} 
for all $a\in C_0(\C^4)$, $v\in H$ and $w\in K$. Evaluation of the coordinate functions $M_j$, $N_j$ at a point of the space ${\sfM}^{k,n}$ yields precisely the self-conjugate monad labelled by that point.

Applying the deformation procedure therefore yields a pair of families of module maps
\begin{align}\label{nc-maps}
\hat\rho_z:H\otimes C_0(\C^4_\sigma)\to \A[{\sf M}^{k,n}_\sigma] \otimes K\otimes C_0(\C^4_\sigma),\qquad \hat\rho_z&: v\otimes a \mapsto \sum\nolimits_{j}
\hat M_j \otimes v \otimes \mathfrak{z}(\hat z_j) a, \\
\hat\tau_z:K\otimes C_0(\C^4_\sigma)\to \A[{\sf M}^{k,n}_\sigma] \otimes L\otimes C_0(\C^4_\sigma), \qquad \hat\tau_z&: w\otimes a \mapsto \sum\nolimits_{j}
\hat N_j \otimes w \otimes \mathfrak{z}(\hat z_j) a, 
\end{align} 
where $a\in C_0(\C^4_\sigma)$, $v\in H$ and $w\in K$. The elements $\hat z_j$, $j=1,\ldots,4$, denote the generators of the coordinate algebra $\A[\C^4_\sigma]$. Moreover, we write $\hat M_j$, $\hat N_l$ for the elements affiliated to $C_0({\sfM}^{k,n}_\sigma)$ obtained by conjugating the matrix elements $M_j$, $N_l$ by appropriate unitaries, in analogy with eq.~\eqref{units}. 

Since the torus action preserves the monad condition $\tau_z\circ\rho_z=0$, we automatically find that $\hat\tau_z\circ\hat\rho_z=0$. We define the coordinate algebra $\A[{\sfM}^{k,n}_\sigma]$ of the noncommutative parameter space ${\sfM}^{k,n}_\sigma$ to be the unital $*$-algebra generated by the conjugated matrix elements $\hat M_j^{ab}$, $\hat N_l^{cd}$, for each $j,l=1,\ldots,4$ and  $a.d=1,\ldots,k$, $b,c=1,\ldots,2k+n$, subject to the relations coming from the monad condition (see \cite{bl:mod} for further details).

Although the parameter space $\mathsf{M}^{k,n}_\sigma$ has fewer classical points than the original space ${\sfM}^{k,n}$, we may nevertheless work with the whole family at once, by considering it as parametrizing a {\em quantum family of maps} in the sense of  \cite{wor1,lprs:ncfi}. Moreover, as shown in \cite{bl:mod,bvs}, there is a noncommutative version of the ADHM construction which produces a family of instantons on the quantum sphere $S^4_\sigma$ parametrized by the space ${\sfM}^{k,n}_\sigma$, that is to say a noncommutative family of Hermitian vector bundles over $S^4_\sigma$ with rank $n$ and second Chern class $k$, equipped with a canonical family of connections having anti-self-dual curvature (since the homology groups of the noncommutative sphere $S^4_\sigma$ are isomorphic to those of the classical four-sphere \cite{blvs}, the first Chern class of this family of bundles is automatically zero). It is beyond the scope of this paper to recall the construction here: we are interested more in the structure of the noncommutative parameter space ${\sfM}^{k,n}_\sigma$ itself.

\subsection{Gauging away noncommutative parameters} In this final section of the paper we investigate how the internal gauge group of the quantum sphere $S^4_\sigma$ interacts with the topology of the parameter space ${\sfM}^{k,n}_\sigma$. In this case the algebra $C(S^4_\sigma)$ is noncommutative and so the gauge theory of instantons on $S^4_\sigma$ is now also affected by the internal gauge group $\textup{U}(A)$ of the function algebra $A=C(S^4_\sigma)$.

Once again we rely on the fact that the $C^*$-algebra $A=C(S^4_\sigma)$ has a spectral decomposition
$$
A=\bigoplus_{\zeta\in\ZZ^2}A_\zeta.
$$
This determines an action of the dual group $\hGamma=\ZZ^2$ upon $A$ by $*$-automorphisms, according to the formula
\begin{equation}\label{zz}
\hat\delta:\hGamma\to \textup{Aut}(A),\qquad \hat\delta_\xi(a):=\sigma^{-2}(\zeta,\xi)\,a,
\end{equation}
defined here for each homogeneous element $a\in A_\zeta$ and extended by linearity. The fact that $\sigma$ is a bicharacter means that this is a well-defined group action. This is a version of the `bosonization' action described by Majid for Hopf algebras \cite{ma:book}, but now at the level of group actions on $C^*$-algebras.

\begin{prop}\label{pr:inn}
The group action $\hat\delta:\hGamma\to \textup{Aut}(A)$ is by inner automorphisms of $A$.
\end{prop}

\proof We consider $\mathpzc{E}:=A$ as a right Hilbert module over itself by right multiplication, with $A$-valued inner product defined by $\la a,b\ra_A:=a^*b$ for each $a,b\in \mathpzc{E}$. We write $\pi:A\to\End^*_A(\mathpzc{E})$ for the representation of $A$ upon $\mathpzc{E}$ by left multiplication.
Then each $\xi\in\hGamma$ gives rise to a representation of $A$ upon $\mathpzc{E}$ by 
$$
\pi_\xi:A\to\mathbb{B}_A(\mathpzc{E}),\qquad \pi_\xi:=\pi\circ\hat\delta_\xi.
$$
For fixed $\xi\in\hGamma$, we get a map 
$$
W_\xi:\mathpzc{E}\to\mathpzc{E},\qquad W_\xi(e)=\sigma^{-2}(\zeta,\xi)\,e,
$$
defined for each homogeneous element $e\in A_\zeta$ and extended by linearity. This defines a unitary isomorphism of Hilbert modules
which intertwines the representations $\pi$ and $\pi_\xi$, whence there exists a unitary element $u\in\End^*_A(\mathpzc{E})$ such that $u\pi(a)u^*=\pi_\xi(a)$ for all $a\in A$ \cite{meyer}. However, since $A$ is unital, we have $\End^*_A(\mathpzc{E})=A$, from which the result follows immediately.\endproof

As a consequence we obtain an injective group homomorphism from the discrete group $\hGamma$ into the internal gauge group of $S^4_\sigma$,
\begin{equation}\label{dischom}
\hat\delta:\hGamma\to \U(A),
\end{equation}
given by sending the element $\xi\in\hGamma$ to the unitary element $u\in\U(A)$ which implements the automorphism $\hat\delta_\xi$. 

The action \eqref{zz} lifts to give a group homomorphism
\begin{equation}\label{lift}
\hat\delta:\hGamma\to\textup{Aut}(C_0(\C^4_\sigma)),\qquad \hat\delta_\xi(a):=\sigma^{-2}(\zeta,\xi)a,
\end{equation}
for each $\xi\in\hGamma$ and each homogeneous element $a\in C_0(\C^4_\sigma)_\zeta$ with respect to the spectral decoposition of $C_0(\C^4_\sigma)$. Given a monad over the noncommutative twistor space, the action of an element $\xi\in\hGamma$ gives rise to a new monad determined by the diagram
$$
\begin{CD} 
M:\qquad @. H\otimes C_0(\C^4_\sigma) @>\rho_z >> K\otimes C_0(\C^4_\sigma) @>\tau_z >> L\otimes C_0(\C^4_\sigma)
\\ @. @VV \id\otimes\hat\delta_\xi V @VV \id\otimes\hat\delta_\xi V @VV \id\otimes\hat\delta_\xi   V\\  M^\xi:\qquad @.  H\otimes C_0(\C^4_\sigma) @>\rho^\xi_z >> K\otimes C_0(\C^4_\sigma) @>\tau^\xi_z >> L\otimes C_0(\C^4_\sigma),
\end{CD}
$$
where the right module maps in the complex $M^\xi$ are defined by
$$
\rho^\xi_z:=(\id\otimes\hat\delta_\xi)\circ\rho_z\circ(\id\otimes\hat\delta_\xi^*),\qquad \tau^\xi_z:=(\id\otimes\hat\delta_\xi)\circ\tau_z\circ(\id\otimes\hat\delta_\xi^*).
$$
A computation along similar lines to those in Lem.~\ref{le:c-map} yields the explicit expressions
$$
\rho^\xi_z:=\sum_j \sigma^{-2}(\zeta_j,\xi)M_j\otimes\hat z_j,\qquad \tau^\xi_z:=\sum_j\sigma^{-2}(\zeta_j,\xi)N_j\otimes \hat z_j,
$$
for the module maps in the monad $M^\xi$. By construction, the instanton bundles corresponding to the monads $M$ and $M^\xi$ via the noncommutative ADHM construction are gauge equivalent under the internal gauge symmetry $\hat\delta_\xi$. Indeed, it is explained in \cite{bl:mod} that the action of $\hGamma$ on monads induces an action upon the corresponding instanton connections via conjugation by a unitary automorphism. This is a purely `quantum' phenomenon which becomes trivial in the classical limit where $\sigma$ is the trivial bicharacter.

More generally, the same idea works for noncommutative families of monads over $\C^4_\sigma$. Once again regarding the matrix elements $\hat M_j^{ab}$ and $\hat N_j^{cd}$ as generators of the coordinate algebra $\A[{\sfM}^{k,n}_\sigma]$, under the gauge transformation $\delta_\xi$ the quantum families of maps \eqref{module maps}--\eqref{modmaps2} become
$$
\hat\rho^\xi_z:=\sum_j \sigma^{-2}(\zeta_j,\xi)\hat M_j\otimes\hat z_j,\qquad \hat\tau^\xi_z:=\sum_j\sigma^{-2}(\zeta_j,\xi)\hat N_j\otimes \hat z_j,
$$
hence defining an action of $\hGamma$ upon $C_0({\sfM}^{k,n}_\sigma)$ by gauge transformations:
$$
\hat\delta:\hGamma\to \textup{Aut}(C_0({\sfM}^{k,n}_\sigma)),\qquad \hat\delta_\xi(a)=\sigma^{-2}(\zeta,\xi)\,a,
$$
for each $j=1,\ldots,4$ and each homogeneous element $a\in C_0({\sfM}^{k,n}_\sigma)_\zeta$, then extended to $C_0({\sfM}^{k,n}_\sigma)$ by linearity. Once again it is shown in \cite{bl:mod} that the corresponding noncommutative families of instantons produced by the noncommutative ADHM construction are related via conjugation by a unitary operator under the action of the group $\hGamma$. 

It follows that the `points' of the noncommutative space ${\sfM}^{k,n}_\sigma$ which lie in the same $\hGamma$-orbit describe gauge equivalent instanton connections. Consequently, we should quotient the space ${\sfM}^{k,n}_\sigma$ by the action of $\hGamma$ in order to obtain a more efficient space of parameters. In this way, we obtain a covariant system $(C_0({\sfM}^{k,n}_\sigma),\hGamma,\hat\delta)$ and hence an associated crossed product $C^*$-algebra $B:=C_0({\sfM}^{k,n}_\sigma)\lcross_{\hat\delta}\,\hGamma$ describing the quotient of the space ${\sfM}^{k,n}_\sigma$ by the gauge transformations induced by the action of the Pontryagin dual group $\hGamma$.

Our goal is to obtain different factorizations of the crossed product algebra $B$ by equipping it with a family of $\hGamma$-product structures. Recall that, in order to have such a structure, we need a pair of group homomorphisms
$$
\Gamma\to\textup{Aut}(B),\qquad \hGamma\to\U\M(B),
$$
satisfying the conditions of Def.~\ref{de:g-prod} (although now in the dual picture with $\Gamma$ and $\hGamma$ interchanged). We begin by taking $\hat\lambda:\hGamma\to\U\M(B)$ to be the canonical inclusion described in eq.~\eqref{canincs} of $\hGamma$ into the crossed product $C_0({\sfM}^{k,n}_\sigma)\lcross_{\hat\delta}\,\hGamma$.

On the other hand, for each $r=(e^{2\ii\pi r_1},e^{2\ii\pi r_2})\in\Gamma$ and each $m=(m_1,m_2)\in \hGamma$, let us define a group homomorphism  
\begin{equation}\label{wind}
\tilde\beta^m:\Gamma\to \textup{Aut}(C_0({\sfM}^{k,n}_\sigma)),\qquad \tilde\beta^m (r):=\tilde\beta\left((e^{2\ii\pi m_1r_1},e^{2\ii\pi m_2r_2})\right),
\end{equation}
where $\tilde\beta:\Gamma\to\textup{Aut}(C_0({\sfM}^{k,n}_\sigma))$ is the torus action \eqref{monact}. We shall refer to the element $m\in \hGamma$ as the {\em winding number} of the homomorphism $\beta^m$ defined in eq.~\eqref{wind}. Each such winding number gives rise to an action of $\Gamma$ on $B=C_0({\sfM}^{k,n}_\sigma)\lcross_{\hat\delta}\,\hGamma$ via the diagonal action, defined by
\begin{equation}\label{m-action}
\alpha^m:\Gamma\to\textup{Aut}(B),\qquad (\alpha^m_r(f))(\xi):=\xi(r)\,\beta^m_r(f(\xi)),
\end{equation}
for each $f\in C_c(\hGamma,C({\sfM}^{k,n}_\sigma))$, $r\in\Gamma$, and extended by continuity.

\begin{lem}
For each $m\in \hGamma$ the datum $(B,\alpha^m,\hat\lambda)$ constitutes a $\hGamma$-product.
\end{lem}

\proof It follows immediately from the definitions that $\alpha^m_r(\hat\lambda_\xi)=\xi(r)\hat\lambda_\xi$ for all $r\in\Gamma$ and all $\xi\in\hGamma$, which is all we needed to check.\endproof

We write $C_0(\mathfrak{M}^{k,n}_m)$ for the Landstad algebra of the crossed product $C^*$-algebra $B$ determined by the $\hGamma$-product $(B,\alpha^m,\hat\lambda)$. The Landstad theory of \S\ref{se:land} immediately implies that, for each winding number $m\in\hGamma$, we have an isomorphism $B\simeq C_0(\mathfrak{M}^{k,n}_m)\lcross\, \hGamma$. In particular, we have $C_0({\sfM}^{k,n}_\sigma)\simeq C_0(\mathfrak{M}^{k,n}_0)$.

For each $j,\xi\in\hGamma$, we introduce the families of distributions
$$
f_{j,\xi}\in\M(B),\qquad f_{j,\xi}(\eta) \in C_0({\sfM}^{k,n}_\sigma)_\xi\qquad  \text{if}~\eta=j
$$
and $f_{j,\xi}(\eta)=0$ otherwise. These are simply the set of distributions concentrated at the point $j\in\hGamma$ taking values in the $\xi$-spectral subspace $C_0({\sfM}^{k,n}_\sigma)_\xi$. The action \eqref{m-action} of $\Gamma$ on $\M(B)$ induces a $\hGamma$-grading upon $\M(B)$ for which the $\zeta$-spectral subspace is given by
\begin{equation}\label{grad}
\M(B)_\zeta=\left\{  f_{\zeta-\xi^m,\xi}\in\M(B)~|~\xi\in \hGamma\right\},
\end{equation}
where we have used the notation $\xi^m:=(m_1\xi_1,m_2\xi_2)$ for each $\xi=(\xi_1,\xi_2)\in\hGamma$.

\begin{prop}
The Landstad algebra $B^m:=C_0(\mathfrak{M}^{k,n}_m)$ for the action \eqref{m-action} is given by
$$
B^m=\bigoplus_{\zeta\in\hGamma}\,\left\{  f_{-\zeta^m,\zeta}\in\M(B)\right\},
$$
whose $\zeta$-spectral subspace is the set of distrubutions concentrated at $-\zeta^m$ and taking values in $C_0({\sfM}^{k,n}_\sigma)_\xi$.
\end{prop}

\proof The fixed points in $\M(B)$ of the $\Gamma$-action \eqref{m-action} are precisely those of the degree zero subalgebra for the grading  \eqref{grad}, which is the algebra $B^m$ as stated. The second and third conditions of Def.~\ref{de:l-cons} are now obvious.\endproof

Let us check that there is a choice of winding number
$m\in\hGamma$ for which the algebra of parameters $C_0(\mathfrak{M}^{k,n}_m)$ is commutative. 

\begin{thm}\label{th:final}
There exists $m\in \hGamma$ and a factorization $B\cong C_0(\mathfrak{M}^{k,n}_m)\lcross\, \hGamma$ such that $C_0(\mathfrak{M}^{k,n}_m)$ is a commutative $C^*$-algebra.
\end{thm}

\proof On the homogeneous functions $f_{j,\zeta}$ and $f_{l,\xi}$, the product \eqref{conv} in the multiplier algebra $\M(B)$ simplifies greatly to
\begin{align*}
(f_{j,\zeta}\star f_{l,\xi})(\eta)&=\int_{\hGamma}f_{j,\zeta}(\gamma)\, \hat\delta_\gamma \left(f_{l,\xi}(\eta-\gamma)\right)\D \gamma \\
&=f_{j,\zeta}(j)\, \hat\delta_j \left(f_{l,\xi}(l)\right)\\ &=\sigma^{-2}(\xi,j)f_{j,\zeta}(j)f_{l,\xi}(l)
\end{align*}
if $\eta=j+l$ and zero otherwise. The relations in the Landstad algebra $C_0(\mathfrak{M}^{k,n}_m)$ are therefore given by
\begin{align*}
f_{-\zeta^m,\zeta}\star f_{-\xi^m,\xi}&=\sigma^{-2}(-\zeta^m,\xi)\,\sigma^{-2}(\xi,\zeta)\,\sigma^{-2}(\zeta,-\xi^m)\, f_{-\xi^m,\xi}\star f_{-\zeta^m,\zeta}. 
\end{align*}
Since $\sigma$ is a group bicharacter, we may expand the coefficients involving $\sigma^{-2}$ in terms of the generators of the discrete group $\hGamma$, effectively reducing the situation to the algebraic case of \cite{bl:mod}. As shown there, any choice of $m=(m_1,m_2)$ for which $m_1+m_2=1$ makes the resulting
algebra $C_0(\mathfrak{M}^{k,n}_m)$ commutative.\endproof

Thus we have shown that, up to gauge transformations induced by the inner automorphisms of the $C^*$-algebra $C(S^4_\sigma)$, the system of instantons described by the space of monads over noncommutative twistor space $\C^4_\sigma$ is determined equally well by each of the quantum parameter spaces $\mathfrak{M}^{k,n}_m$. Moreover, by making a certain choice of internal gauge, this parameter space may always be chosen to be classical.

Within the parameter spaces $\mathfrak{M}^{k,n}_m$ there still remains the gauge freedom afforded by the finite dimensional group $G$ of eq.\eqref{fd-group}. Indeed, recall that, in order to compute the moduli space of instantons on the classical sphere $S^4$, it is sufficient to replace the infinite dimensional gauge group by its finite dimensional counterpart. Moreover, in the above computations, we have replaced the infinite dimensional group $\U(\A)$ by the discrete group $\hGamma=\ZZ^2$. It is therefore natural to conjecture that, in order to compute the moduli space of instantons on the quantum sphere $S^4_\sigma$, it is sufficient to replace the infinite dimensional gauge groups $\mathcal{G}_e(\mathpzc{E})$ and $\U(A)$ by the much simpler groups $G$ and $\hGamma$.

\section{Concluding Remarks}

We conclude with a summary of the present paper, together with a few remarks regarding some technical points which have been overlooked and some possible directions of future research.

As we have seen, the action of the group $\Gamma=\TT^2$ upon the $C^*$-algebra $C_0(\C^4)$ automatically induces an action of $\Gamma$ on the space of monads over twistor space $\C^4$. The functorial nature of the Landstad--Kasprzak deformation procedure means that, in passing from $C_0(\C^4)$ to the noncommutative algebra $C_0(\C^4_\sigma)$, we cannot avoid deforming the space ${\sfM}^{k,n}$ into the `quantum' family of monads ${\sfM}^{k,n}_\sigma$ over noncommutative twistor space $\C^4_\sigma$.

The points of the space ${\sfM}^{k,n}$ corresponding to $\Gamma$-equivariant monads (that is to say, monads with the property that $M\sim M^r$ for all $r\in \Gamma$) survive as classical points of the noncommutative space ${\sfM}^{k,n}_\sigma$. This is equivalent to saying that the $\Gamma$-equivariant instantons on the classical sphere $S^4$ are preserved by the deformation procedure and automatically yield instantons on the quantum sphere $S^4_\sigma$ (see \cite{blvs} for further details). On the other hand, the points of ${\sfM}^{k,n}$ which correspond to monads which are not $\Gamma$-equivariant become `quantized' by the deformation procedure. Nevertheless, Thm.~\ref{th:final} shows that every equivalence class for the action of the subgroup $\hGamma$ of the group of internal gauge transformations of $S^4_\sigma$ contains a classical point.

In this paper we do not claim that all instantons on $S^4_\sigma$ arise in this way from a monad construction, although the results of \cite{blvs} seem to indicate that this is the case. However, even in the noncommutative case, equivalence of monads in the sense of Def.~\ref{de:eq-mon} under the action of the group \eqref{fd-group} continues to imply gauge equivalence of the corresponding instantons ({\em cf}. \cite{bvs} for further discussion in this direction).

As a way to interpret the noncommutative geometry of Thm.~\ref{th:final}, we recall that the correct interpretation of quotient spaces in noncommutative geometry is via crossed product $C^*$-algebras \cite{ac:book}. This means that (up to Morita equivalence) the $C^*$-algebra $C_0(\mathfrak{M}^{k,n}_m)\lcross\, \hGamma$ should be thought of as the algebra of continuous functions on the quotient space $\mathfrak{M}^{k,n}_m/\hGamma$. The fact that, for each $m\in\hGamma$, these crossed product $C^*$-algebras are all isomorphic leads to the following diagram of noncommutative spaces:
\begin{equation*}
\xymatrix{ \mathsf{M}^{k,n}_\sigma \ar[d]  & & \ar[d] {\mathfrak{M}^{k,n}_m} \\
{\mathsf{M}^{k,n}_\sigma/\hGamma}  & \simeq & {\mathfrak{M}^{k,n}_m/\hGamma}}
\end{equation*}
As a consequence, we are directed towards a `stacky' interpretation of the moduli space of instantons on $S^4_\sigma$. The noncommutative spaces ${\mathfrak{M}^{k,n}_m}$ are all equivalent presentations of the `stack' ${\mathsf{M}^{k,n}_\sigma/\hGamma}$. Since in noncommutative geometry there is no reason to prefer classical spaces over noncommutative ones, these presentations should be thought of as equivalent parameter spaces for the same system of instantons. It remains to be seen how one should define a topological stack over an appropriate category of noncommutative spaces, in order to make this idea precise.

A small but important technical point we have overlooked (since it did not play a particularly important role in the present paper) is that, strictly speaking, the space ${\sf M}^{k,n}$ of self-conjugate monads introduced in Def.~\ref{de:scmon} is not parametrized by the matrix elements $M_j^{ab}$ and $N_j^{cd}$, as claimed in \S\ref{se:cl-adhm}. 
Indeed, these matrix elements parametrize the space of all self-conjugate pairs of module maps $\rho_z$ and $\tau_z$ such that $\tau_z\circ\rho_z=0$, with no guarantee that $\rho_z$ is injective and $\tau_z$ is surjective, as demanded by eq.~\ref{monad}. The algebra $\A[{\sf M}^{k,n}]$ is therefore rather the coordinate algebra of a completion of the space of monads obtained by relaxing these `non-degeneracy' conditions ({\em cf} .\cite{dk}). An important point to be understood is the way in which this completion of the space of monads on classical twistor space $\C^4$ is related to the corresponding completion of the space of monads on noncommutative twistor space $\C^4_\sigma$.

\end{document}